\documentclass[12pt]{iopart}

\usepackage{iopams}  
\usepackage{graphicx}

\usepackage[square,sort&compress,numbers]{natbib}

\begin{document}

\title{Revisiting time delay interferometry for unequal-arm LISA and TAIJI}

\author{Gang Wang}
\ead{gwang@shao.ac.cn, gwanggw@gmail.com}
\address{Shanghai Astronomical Observatory, Chinese Academy of Sciences, Shanghai 200030, China}

\author{Wei-Tou Ni}
\ead{wei-tou.ni@wipm.ac.cn, weitou@gmail.com}
\address{State Key Laboratory of Magnetic Resonance and Atomic and Molecular Physics, Innovation Academy for Precision Measurement Science and Technology (APM), Chinese Academy of Sciences, Wuhan 430071, China}
\address{Department of Physics, National Tsing Hua University, Hsinchu, Taiwan, 30013, ROC}

\vspace{10pt}
\begin{indented}
\item[]
\end{indented}

\begin{abstract}

Three spacecraft of LISA/TAIJI mission follow their respective geodesic trajectories, and their interferometric arms are unequal and time-varying due to orbital dynamics. Time-delay interferometry (TDI) is proposed to suppress the laser frequency noise caused by the unequal-arm. By employing the numerical orbit, we investigate the sensitivity of the first-generation TDI configurations and their corresponding optimal A, E, and T channels. The sensitivities of T channels from Michelson and Monitor/Beacon configurations diverge from the equal-arm case in frequencies lower than 10 mHz, and their performances vary with the inequality of the arm lengths. The mismatches of the laser beam paths are evaluated in a dynamic case, and the residual laser noise in the first-generation TDI could not satisfy the mission requirement.

\end{abstract}

\maketitle

\section{Introduction}

Gravitational wave (GW) started to become a new method to observe the universe since the first detection of advanced LIGO -- GW150914 \cite{Abbott:2016blz}.
During the advanced LIGO and advanced Virgo O1 -- O3 run, GW signals from stellar-mass compact binary coalescences were frequently detected \cite[and references therein]{TheLIGOScientific:2016pea,TheLIGOScientific:2017qsa,LIGOScientific:2018mvr,Nitz:2019hdf,Abbott:2020uma,LIGOScientific:2020stg,Abbott:2020khf,gracedb}, and were utilized to study the fundamental physics and astronomy \cite[and references therein]{Abbott:2017xzu,Soares-Santos:2019irc,Abbott:2019yzh,TheLIGOScientific:2016src,Abbott:2018lct,LIGOScientific:2019fpa}.
With the joining of KAGRA \cite{Akutsu:2020his,Michimura:2020xnj}, the detectability and sky localization for the GW signals will be improved by the LIGO-Virgo-KAGRA network \cite{Aasi:2013wya}.

Besides the high-frequency GW (10--2000 Hz) observed by the ground-based interferometers, the research and development are also thriving in other frequency bands. In the middle frequency band (0.1--10 Hz), various ground-based and space-borne GW detectors have been proposed. The ground-based detectors include Atom Interferometer (MIGA \cite{Geiger:2015,Junca:2019}, MAGIS-100 \cite{Coleman:2018}, ZAIGA \cite{Zhan:2019}, ELGAR \cite{Canuel:2019} and AION \cite{Badurina:2020} ), MI (Michelson Interferometer) \cite{2013PhRvD..88l2003H}, SOGRO (Superconducting Omni-directional Gravitational Radiation Observatory) \cite{Paik:2016,Paik:2019}, and TOBA (Torsion-Bar Antenna) \cite{Shimoda:2019}. 
In space, by employing different approaches and/or configurations, multiple missions have been proposed: BBO \cite{Crowder:2005nr} and DECIGO \cite{Kawamura:2006up,Kawamura:2018esd}, AEDGE \cite{Abou:2019}, AIGSO \cite{Gao:2017rgh,Wang:2019aigso}, AMIGO \cite{Ni:2018,Ni:2019amigo}, B-DECIGO \cite{Kawamura:2018esd}, DO \cite{Sedda:2019}, and INO \cite{Ebisuzaki:2020}.

In the milli-Hz frequency band (0.1 mHz--1 Hz), besides LISA \cite{LISA2000,2017arXiv170200786A}, two Chinese space missions were proposed--TAIJI \cite{Hu:2017mde} and TianQin \cite{Luo:2015ght}. The TAIJI mission is considered to be a LISA-like configuration in a heliocentric orbit leading the Earth by 20$^\circ$, and TianQin uses triangular interferometry in a geocentric orbit configuration. The studies about TAIJI and TianQin are actively ongoing \cite[and references therein]{Wu:2019thj,Wang:2020a,2019arXiv190701838W,2019PhRvD.100f4033Z,Luo:2020tianqin,Luo:2020taiji,Ye:2019txh,Luo:2020}. Beyond the LISA, larger space-borne interferometers were proposed to observe GW in the $\mu$Hz frequency band including ASTROD-GW \cite{Ni:2013}, $\mu$-Aries \cite{Sesana:2019}, Folkner's mission \cite{Folkner:2019}, and Super-ASTROD \cite{Ni:2009}.

For space missions, with state-of-the-art technology, the laser frequency noise is still too overwhelming to observe GW. Time-delay interferometry (TDI) is proposed for LISA-like missions to suppress the laser frequency noise. The previous studies have demonstrated that the TDI could effectively suppress laser noise \cite[and references therein]{1999ApJ...527..814A,2000PhRvD..62d2002E,2001CQGra..18.4059A,2003PhRvD..67l2003T,Tinto:2003vj,Shaddock:2003dj,Vallisneri:2004bn,2008PhRvD..77b3002P,Dhurandhar:2010pd,Tinto:2014lxa,2019PhRvD..99h4023B}. On the other side, the GW response of TDI could also be suppressed by combining the time-shifted measurements.
To investigate the performances of TDI, multiple simulators were developed for the LISA mission. The {\it LISA Simulator} was developed by the group at Montana State University to evaluate the response function and noise \cite{2003PhRvD..67b2001C,2004PhRvD..69h2003R}. The {\it Synthetic LISA} was developed by Vallisneri to simulate the LISA measurement process considering the level of scientific and technical requirements \cite{Vallisneri:2004bn}. {\it LISACode} was built to pave the road between the basic principles of LISA and sophisticated simulator \cite{2008PhRvD..77b3002P}, and its successor {\it LISANode} is developed to adapt to the new LISA design \cite{2019PhRvD..99h4023B}. For the unequal-arm scenario, Larson \etal \cite{Larson:2002xr} analytically examined the impacts of the unequal arm on the transfer function and the sensitivity for a LISA-like mission.

In our previous works \cite{Wang:2020a,Wang:2020cpq}, by using the numerical orbit, we evaluated detectabilities of the first-generation and second-generation TDI channels. 
In this paper, we focus on the sensitivity investigations of the optimal channels (A, E, and T) from the first-generation TDI channels \cite{Prince:2002hp,Vallisneri:2007xa}. By considering the time-varying unequal-arm numerical orbits and three noise sources (laser frequency noise, acceleration noise, and optical metrology noise), we evaluate the response functions and the noise levels in the optimal channels. The results show that performances of the T channels from the Michelson and Monitor configurations are divergent from the equal-arm case and are sensitive to the inequality of arm lengths. Considering residual laser noise is related to the mismatch of interferometric paths, we numerically calculate the mismatches of laser paths in TDI channels. The acceleration noise and optical metrology noise is formulated by distinguishing the different arms. The average sensitivities are synthesized by weighting the noise PSD with the GW response. 

This paper is organized as follows. 
In Sec. \ref{sec:recipe}, we introduce the recipe in our investigation including the numerical mission orbits, TDI optimal channels, the GW response function, and noise assumptions for sensitivity evaluations. 
In Sec. \ref{sec:Michelson_result}, we specify the sensitivity evaluations for the optimal channels from the Michelson TDI configuration, and we formulate the Michelson T channel performances with the inequality of the arms.
The optimal channels from other first-generation TDI channels are examined in Sec. \ref{sec:other_TDI_configurations}.
And we recapitulate our conclusions in Sec. \ref{sec:conclusions}.
(We set $G=c=1$ in this work.)

\section{Recipe of the investigations} \label{sec:recipe}

\subsection{Numerical mission orbits}

The LISA is proposed to be $2.5 \times 10^6$ km interferometric arms and trailing the Earth by $20^\circ$, and the constellation formed by three S/C is $60^\circ$ inclined with respect to the ecliptic plane \cite{2017arXiv170200786A}. The TAIJI is planned to be a LISA-like mission with $3 \times 10^6$ km arms and leading (or trailing) the Earth by $20^\circ$ \cite{Hu:2017mde,Luo:2020}.

\begin{figure*}[htb]
\includegraphics[width=0.48\textwidth]{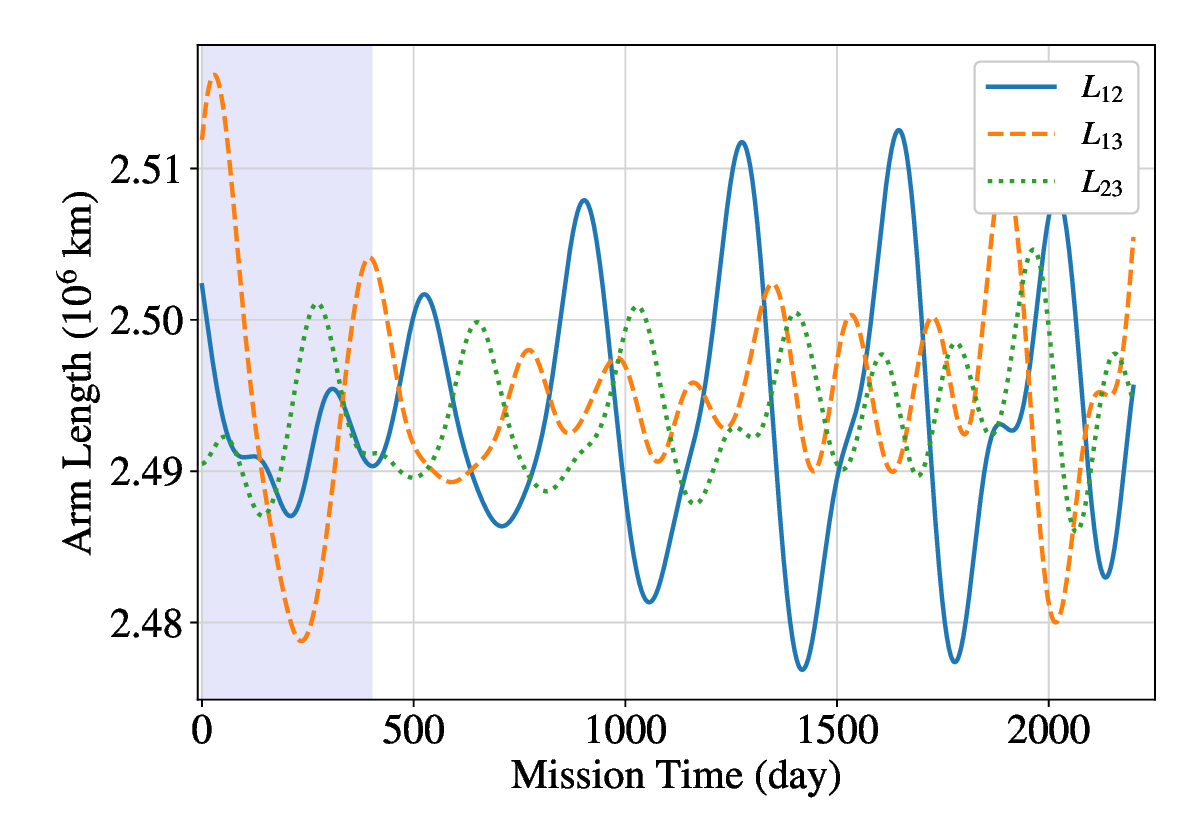}
\includegraphics[width=0.48\textwidth]{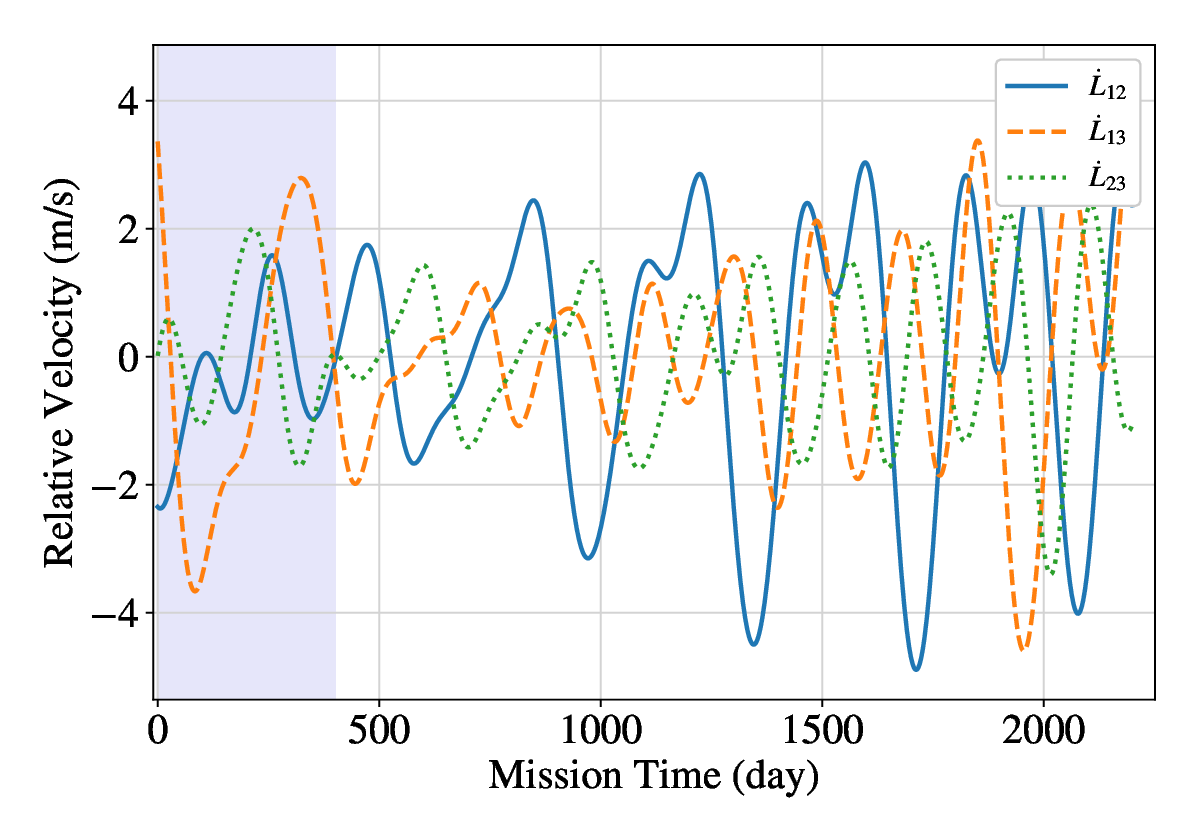}
\includegraphics[width=0.48\textwidth]{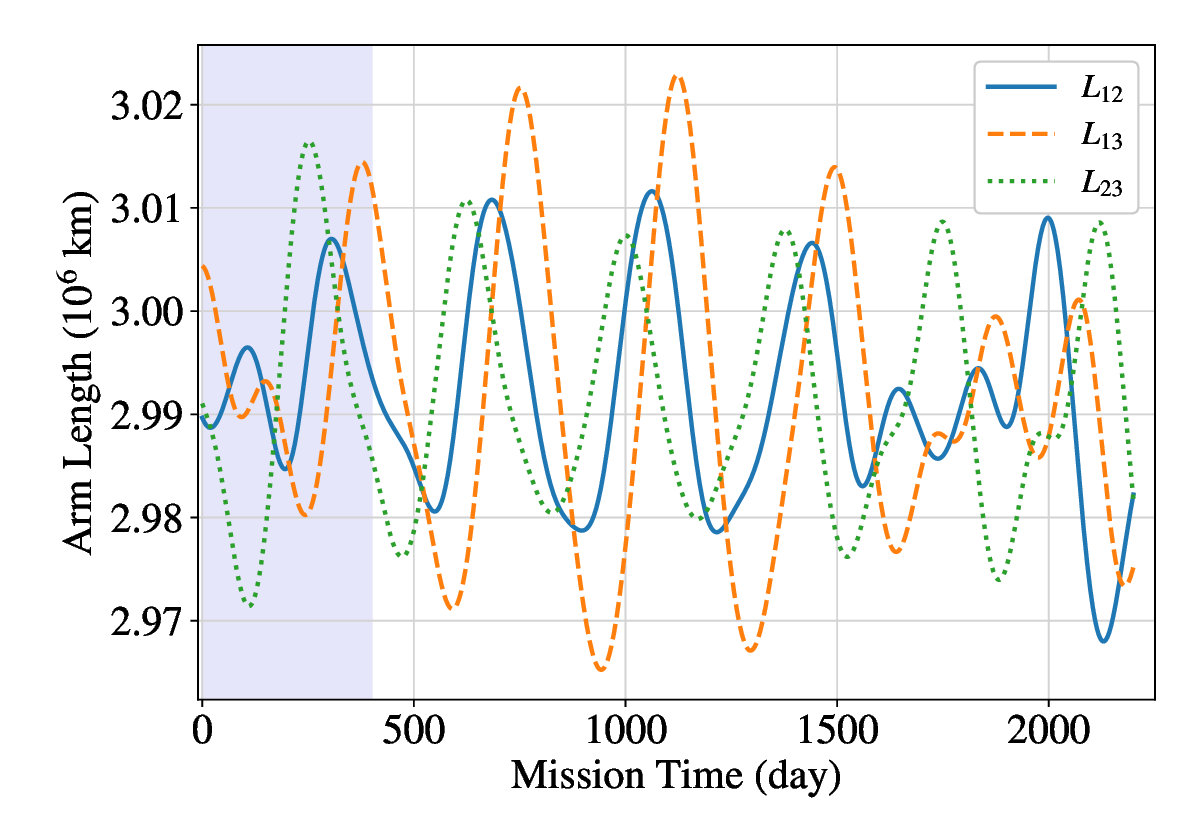}
\includegraphics[width=0.48\textwidth]{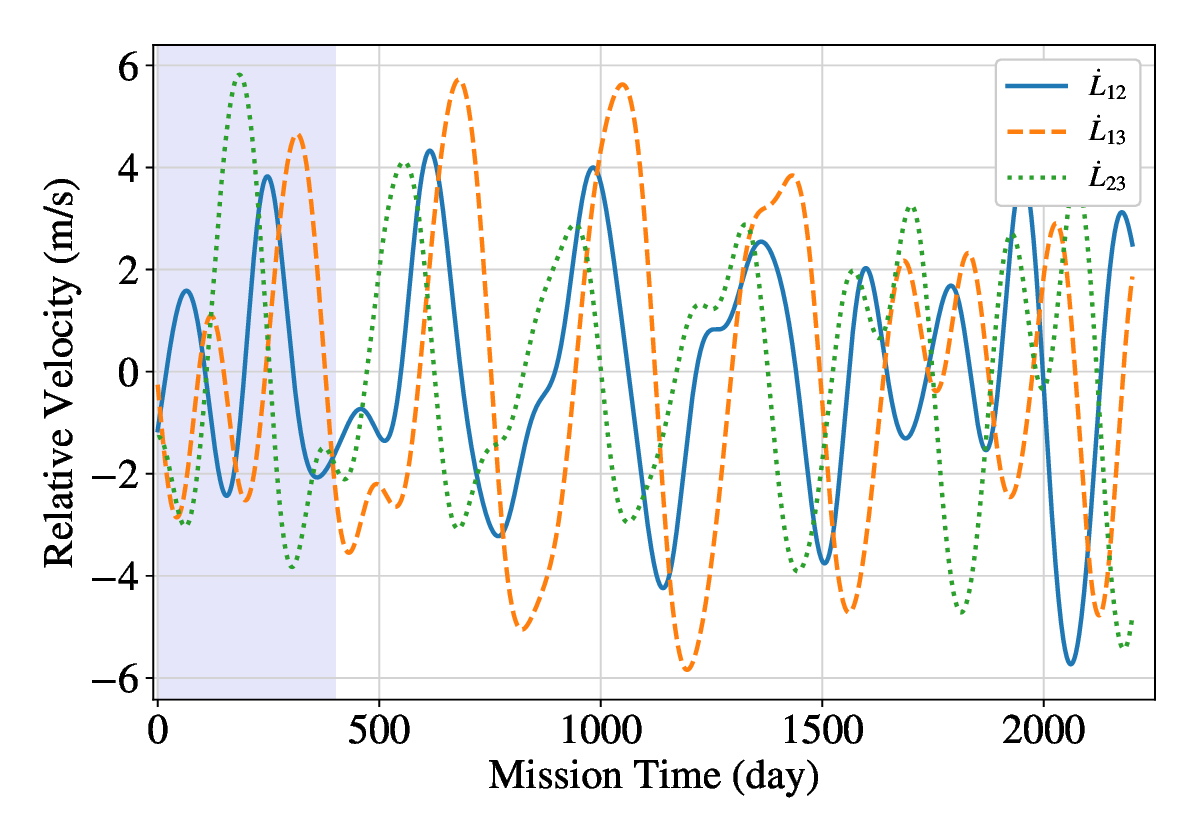}
\caption{\label{fig:orbit} The numerical orbits for the LISA (upper row) and TAIJI (lower row) missions used in our investigations. The arm length $L_{ij}$ changes with time are shown in the left plots, and the relative velocities $\dot{L}_{ij}$ between S/C are shown in the right panel. In the following calculation, the first 400 days shown by the shadow areas are employed to evaluated the performances of TDI channels.}
\end{figure*}

In previous works \cite{Wang:2011,Wang:2014aea,Wang:2012ce,Wang:2012te,Dhurandhar:2011ik,Wang:2014cla,Wang:2017aqq,Wang:2019ipi}, we developed a workflow to design and optimize the orbits for GW space missions by using an ephemeris framework in the solar system barycentric (SSB) coordinates, as well as to calculate the path mismatches between the TDI laser beams. Based on the orbital requirements for new LISA \cite{2017arXiv170200786A}, we achieved the numerical orbit for 6 years satisfying the criteria: 1) the relative velocities between S/C are smaller than 5 m/s; 2) the changes of breathing angles are less than 1 deg, and 3) the trailing angle is in a range [$19^\circ$, $23^\circ$] \cite{Wang:2017aqq}. For the TAIJI mission with longer arms, the relative velocities are loosened up to be less than 6 m/s. The optimized orbits achieved for LISA and TAIJI missions are shown in Figure \ref{fig:orbit}. We select the first 400 days to investigate the sensitivities of TDI channels in the unequal-arm scenarios as shown by the shadow areas.

\subsection{TDI configurations and the optimal channels}

TDI is essential for LISA-like missions to suppress the laser frequency noise and achieve target sensitivity. The principle of TDI is to properly time shift and combine the measurements to form a (nearly) equivalent equal-arm interferometry. 
The first-generation TDI combinations could cancel out the laser frequency noise in a static unequal-arm configuration, and the second-generation TDI combinations could further cancel the frequency noise in a configuration with relative movement. In \cite{Wang:2020cpq}, we have investigated the sensitivities of various second-generation TDI combinations by using a new numerical algorithm. In this work, we focus on the first-generation TDI configuration on the unequal arm case.

\begin{figure*}[thb]
\includegraphics[width=0.24\textwidth]{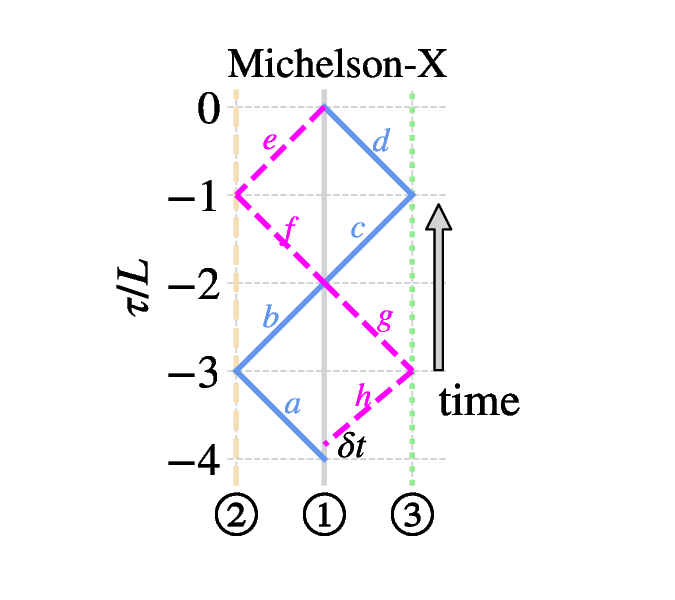}
\includegraphics[width=0.24\textwidth]{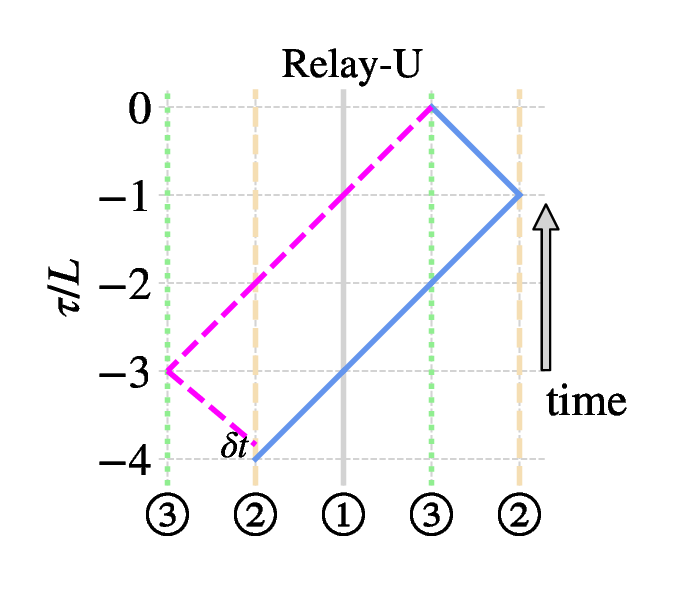}
\includegraphics[width=0.24\textwidth]{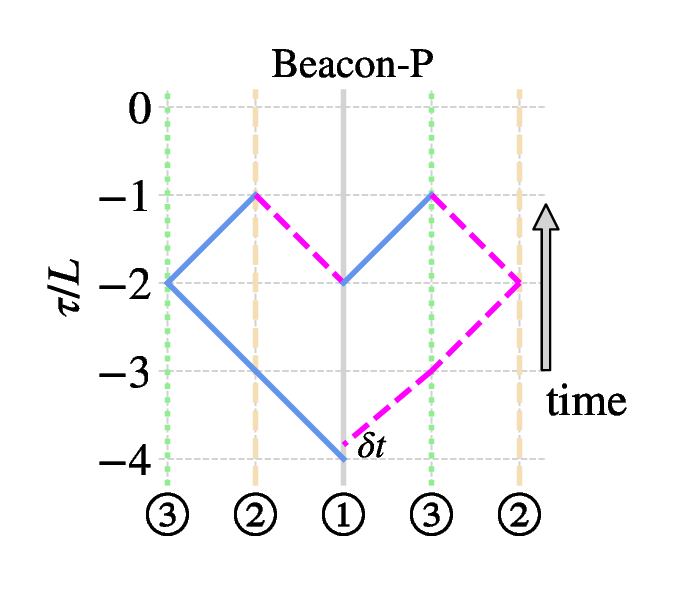}
\includegraphics[width=0.24\textwidth]{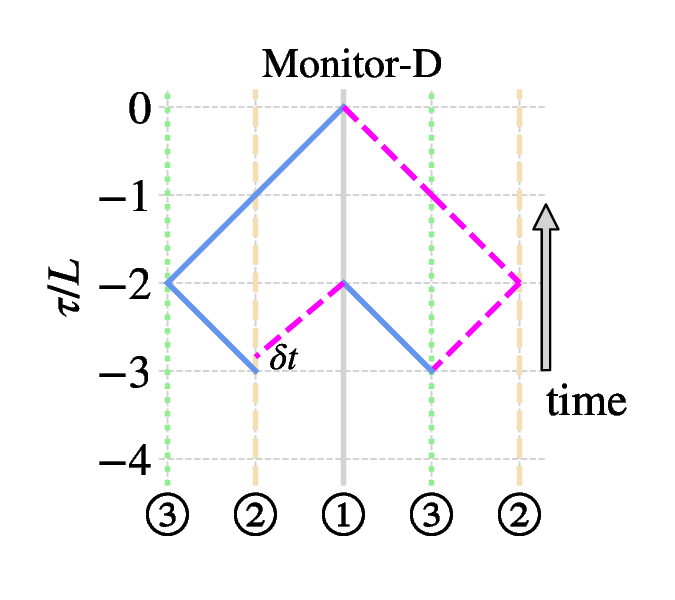}
\caption{\label{fig:X_U_P_D_diagram} The S/C layout-time delay diagrams for Michelson-X, Relay-U, Beacon-P and Monitor-D channels \cite{Wang:2020cpq}. The links $a$-$h$ in Michelson-X plot correspond to the terms in \Eref{eq:X_measurement}. The $\delta t$ indicates the mismatch of the combined beam paths and will be utilized for estimating the residual laser noise. }
\end{figure*}

Five first-generation TDI configurations were developed for the LISA mission which are, Michelson (X, Y, Z), Sagnac ($\alpha$, $\beta$, $\gamma$), Relay (U, V, W), Beacon (P, Q, R), and Monitor (D, F, G) \cite[and references therein]{1999ApJ...527..814A,2000PhRvD..62d2002E,2001CQGra..18.4059A,2003PhRvD..67l2003T,Vallisneri:2005ji,Tinto:2014lxa,Vallisneri:2007xa}. Except for the Sagnac, the first channels from other four TDI configurations are shown by S/C layout-time delay diagrams in Figure \ref{fig:X_U_P_D_diagram} \cite{Wang:2020cpq}. The Michelson-X channel S/C layout-time delay diagram is shown by the first plot, and the links $a$-$h$ correspond to the terms of measurements in the X channel as, 
\begin{equation} \label{eq:X_measurement}
\eqalign{
\fl \qquad \qquad  \mathrm{link }~ a \qquad \quad \mathrm{link } ~ b \qquad \ \ \mathrm{link }  ~c \quad \mathrm{link } ~ d \nonumber  \\
\fl {\rm X} = [ \overbrace{ \mathcal{D}_{31} \mathcal{D}_{13} \mathcal{D}_{21} \eta_{12}  } + \overbrace{ \mathcal{D}_{31}  \mathcal{D}_{13} \eta_{21} } +   \overbrace{ \mathcal{D}_{31} \eta_{13} } + \overbrace{ \eta_{31} }   ] \nonumber  
  - [ \underbrace{ \eta_{21} } + \underbrace{ \mathcal{D}_{21} \eta_{12} } + \underbrace{ \mathcal{D}_{21} \mathcal{D}_{12} \eta_{31} } + \underbrace{ \mathcal{D}_{21} \mathcal{D}_{12} \mathcal{D}_{31} \eta_{13} } ], \\
\ \qquad \qquad \qquad \qquad  \qquad \qquad \qquad \quad \mathrm{link } ~ e \quad \mathrm{link } ~ f \quad \ \ \mathrm{link } ~ g \qquad \qquad  \mathrm{link }~ h  
}
\end{equation}
where $\mathcal{D}_{ij}$ is a time-delay operators and act on a measurement $\eta(t)$ by $ \mathcal{D}_{ij} \eta(t) = \eta(t - L_{ij} ) $. The combined observables $\eta_{ji}$ for S/C$j$ to S/C$i$ ($j \rightarrow i$) denote the designed measurements in \cite{Otto:2012dk,Otto:2015,Tinto:2018kij} and specified in \ref{sec:eta_notations}. 

Three optimal channels, (A, E, and T), can be formed from linear combinations of three regular channels (a, b, c) for each configuration \cite{Prince:2002hp,Vallisneri:2007xa},
\begin{equation} \label{eq:optimalTDI}
 {\rm A} =  \frac{  \mathrm{c} -\mathrm{ a } }{\sqrt{2}} , 
 \quad {\rm E} = \frac{ \mathrm{a} - 2 \mathrm{b} + \mathrm{c} }{\sqrt{6}} , 
 \quad {\rm T} = \frac{ \mathrm{a} + \mathrm{b} + \mathrm{c} }{\sqrt{3}}.
\end{equation}
The b and c channels could be obtained from cyclical permutation of the spacecraft indexes in a channel.
The optimal channels could be treated as the virtual interferometers from the topology approach. Freise \etal \cite{Freise:2008dk} formulated the general expression of response for an interferometer with a rotated angle,
\begin{eqnarray}
h(\kappa) =& \sin \zeta \left[ \left( K_1 \sin 2 \kappa + K_2 \cos 2 \kappa \right) h_{+} + \left( K_3 \sin 2 \kappa + K_4 \cos 2 \kappa \right) h_{\times} \right]
\end{eqnarray}
where $\zeta$ is the opening angle of an interferometer, $\kappa$ is the rotated angle of one interferometer with respect to baseline, $K_n$ are the functions depending on the remaining parameters. Without loss of generality, the three channels of Michelson configuration are labeled as $\mathrm{X}=h(0^\circ),~\mathrm{Y}=h(240^\circ),~\mathrm{Z}=h(120^\circ)$. And then the equivalent interferometers of the optimal channels will be
\begin{eqnarray}
\mathrm{A}_\mathrm{Michelson} = &  \frac{ h(120^\circ) - h(0^\circ) }{\sqrt{2}} = \sqrt{ \frac{3}{2} } h(105^\circ),  \label{eq:equivalent_A} \\ 
\mathrm{E}_\mathrm{Michelson} = & \frac{ h(0^\circ) - 2 h(240^\circ) + h(120^\circ) }{\sqrt{6}} = \sqrt{ \frac{3}{2} } h(150^\circ),  \label{eq:equivalent_E} \\ 
\mathrm{T}_\mathrm{Michelson} = &  \frac{ h(0^\circ) + h(240^\circ) + h(120^\circ) }{\sqrt{3}} = 0.
\end{eqnarray}
The virtual interferometers indicate that the A and E channels are equivalent to a regular channel enlarged by $ \sqrt{ \frac{3}{2}} $ and rotated by $45^\circ$ with respect to each other. And the T channel could be a null data stream to characterize the detector noise. However, as we will see in the following section, the performances of T channels in the unequal-arm case will be divergent from the equal-arm case at low frequencies for the Michelson and Monitor/Beacon configurations.

\subsection{GW response of interferometric link} \label{secsub:TDI_response}

The antenna pattern of a TDI channel is combined with GW responses of time-shifted Doppler measurements in six links. The response function of a single link was formulated in \cite{1975GReGr...6..439E,1987GReGr..19.1101W}. The formulas for TDI are specified in \cite{Vallisneri:2007xa,Vallisneri:2012np}.

For a GW signal from the direction $(\lambda, \beta)$ in the SSB coordinates, where $\lambda $ and $\beta$ are the ecliptic longitude and latitude, the propagation vector will be
\begin{equation} \label{eq:source_vec}
 \hat{k}  = -( \cos \lambda \cos \beta, \sin \lambda \cos \beta ,  \sin \beta ).
\end{equation} 
The plus and cross polarization tensors of the GW signal are
\begin{equation}
\eqalign{
{\rm e}_{+} & \equiv \mathcal{O}_1 \cdot 
\left(\begin{array}{*{10}{c}}
1 & 0 & 0 \\
0 & -1 & 0 \\
0 & 0 & 0
\end{array}\right)
\cdot \mathcal{O}^T_1 \times \frac{1+\cos^2 \iota}{2} , 
\cr
{\rm e}_{\times} & \equiv \mathcal{O}_1 \cdot 
\left(\begin{array}{*{10}{c}}
0 & 1 & 0\\
1 & 0 & 0 \\
0 & 0 & 0
\end{array}\right)
\cdot \mathcal{O}^T_1 \times i (- \cos \iota ), 
}
\end{equation}
with
\begin{equation}
\fl \mathcal{O}_1 =
\left(\begin{array}{*{10}{c}}
\sin \lambda \cos \psi - \cos \lambda \sin \beta \sin \psi & -\sin \lambda \sin \psi - \cos \lambda \sin \beta \cos \psi & -\cos \lambda \cos \beta  \cr
     -\cos \lambda \cos \psi - \sin \lambda \sin \beta \sin \psi & \cos \lambda \sin \psi - \sin \lambda \sin \beta \cos \psi & -\sin \lambda \cos \beta  \cr
         \cos \beta \sin \psi & \cos \beta \cos \psi & -\sin \beta  \cr
\end{array}\right),
\end{equation}
where $\psi$ is the polarization angle. The response to the GW in the link from S/C$i$ to $j$ is
\begin{eqnarray} \label{eq:y_ij}
y^{h}_{ij} (f) =&  \frac{ \sum \hat{n}_{ij} \cdot {\mathrm{ e_p}} \cdot \hat{n}_{ij} }{2 (1 - \hat{n}_{ij} \cdot \hat{k} ) } 
 \times \left[  \exp( 2 \pi i f (L_{ij} + \hat{k} \cdot p_i ) ) -  \exp( 2 \pi i f  \hat{k} \cdot p_j )  \right] ,
\end{eqnarray}
where $\hat{n}_{ij}$ is the unit vector from SC$i$ to $j$, $L_{ij}$ is the arm length from SC$i$ to $j$, $p_i$ is the position of the S/C$i$ in the SSB coordinates, and $\iota$ is the inclination angle of the GW source from the line of sight.

\subsection{Noise budgets}

Laser frequency noise is dominant for LISA-like missions, and it could be effectively suppressed by TDI which constructs an equivalent equal path for static unequal-arm triangular formation \cite[and references therein]{1999ApJ...527..814A,2000PhRvD..62d2002E,2001CQGra..18.4059A,2003PhRvD..67l2003T,Vallisneri:2004bn}. However, for the dynamic case, the residual laser noise would be not sufficiently suppressed by the first-generation TDI channels. 

By assuming the path mismatch of a TDI channel is $\delta t$, the PSD of residual laser frequency noise is expected to be \cite{Vallisneri:2005ji},
\begin{equation} \label{eq:dC}
  S_\mathrm{laser} = |\delta \tilde{C}(f)|^2 \simeq |  2 \pi f \delta t \tilde{C}(f) |^2,
\end{equation}
where $\tilde{C}(f)$ is the one-sided square-root spectrum density of laser source stability. The laser noise is expected to be white and has a one-sided (square-root) spectral density of $30\ {\rm Hz}/\sqrt{\rm Hz}$ corresponding to the power spectrum density $|\tilde{C}(f)| \simeq 1 \times 10^{-13} \ {\rm Hz}^{-1/2}$. The PSD of laser noise is proportional to the square of time difference $\delta t^2$ which indicates the impact of the path inequality from the TDI combination. Considering the residual laser noises is depended on different laser sources, the laser noises in optimal channels are expected to be 
\begin{eqnarray} \label{eq:AET_laser_noise}
S_\mathrm{A,laser} = & \frac{ | \delta \tilde{C}_{\rm c}|^2 + | \delta \tilde{C}_{\rm a} |^2 }{2}, \nonumber \\
S_\mathrm{E,laser} = & \frac{ | \delta \tilde{C}_{\rm a}|^2 + 4 | \delta \tilde{C}_{\rm b} |^2 + | \delta \tilde{C}_{\rm c}|^2}{6}, \\
S_\mathrm{T,laser} = & \frac{  | \delta \tilde{C}_{\rm a}|^2 + | \delta \tilde{C}^2_{\rm b}|^2 + | \delta \tilde{C}_{\rm c}|^2 }{3}. \nonumber
\end{eqnarray}

The requirements of acceleration noise $S_{\rm acc}$ for LISA and TAIJI missions are designed to be the same \cite{2017arXiv170200786A,Luo:2020},
\begin{equation}
 S^{1/2}_{\rm acc} \leq 3 \times 10^{-15} \frac{\rm m/s^2}{\sqrt{\rm Hz}} \sqrt{1 + \left(\frac{0.4 {\rm mHz}}{f} \right)^2 }  \sqrt{1 + \left(\frac{f}{8 {\rm mHz}} \right)^4 }.
\end{equation}
The optical metrology noise $S_{\rm op}$ requirements for LISA and TAIJI missions are slightly different,
\begin{eqnarray}
 S^{1/2}_{\rm op, LISA} & \leq 10 \times 10^{-12} \frac{\rm m}{\sqrt{\rm Hz}} \sqrt{1 + \left(\frac{2 {\rm mHz}}{f} \right)^4 },  \\
S^{1/2}_{\rm op, TAIJI} & \leq 8 \times 10^{-12} \frac{\rm m}{\sqrt{\rm Hz}} \sqrt{1 + \left(\frac{2 {\rm mHz}}{f} \right)^4 }.
\end{eqnarray}
The noise PSDs are estimated by assuming there is no correlation between the different test masses and optical benches.

\section{Michelson TDI configuration and its optimal channels} \label{sec:Michelson_result}

The Michelson TDI configuration is the fiducial case, and we examine its performance for the unequal-arm scenario in this section. To understand the impacts of different factors, we split the investigations into three steps, 1) the response of TDI channels to GW signal, 2) the PSD of noises (including laser frequency noise, optical noise, and acceleration noise) in TDI, and 3) the average sensitivities synthesized from the response and noise levels. 

\subsection{GW Response of Michelson TDI} \label{sec:response}

\begin{figure}[htb]
\includegraphics[width=0.54\textwidth]{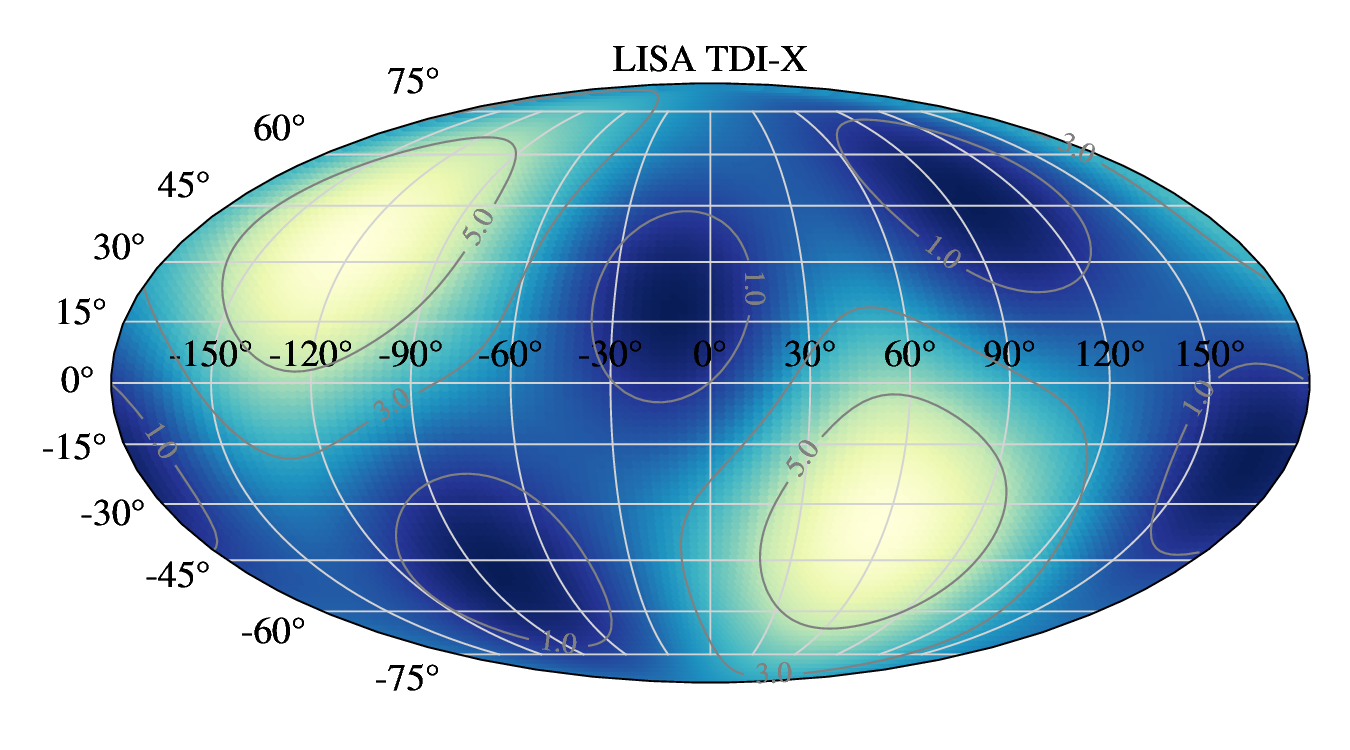}
\includegraphics[width=0.44\textwidth]{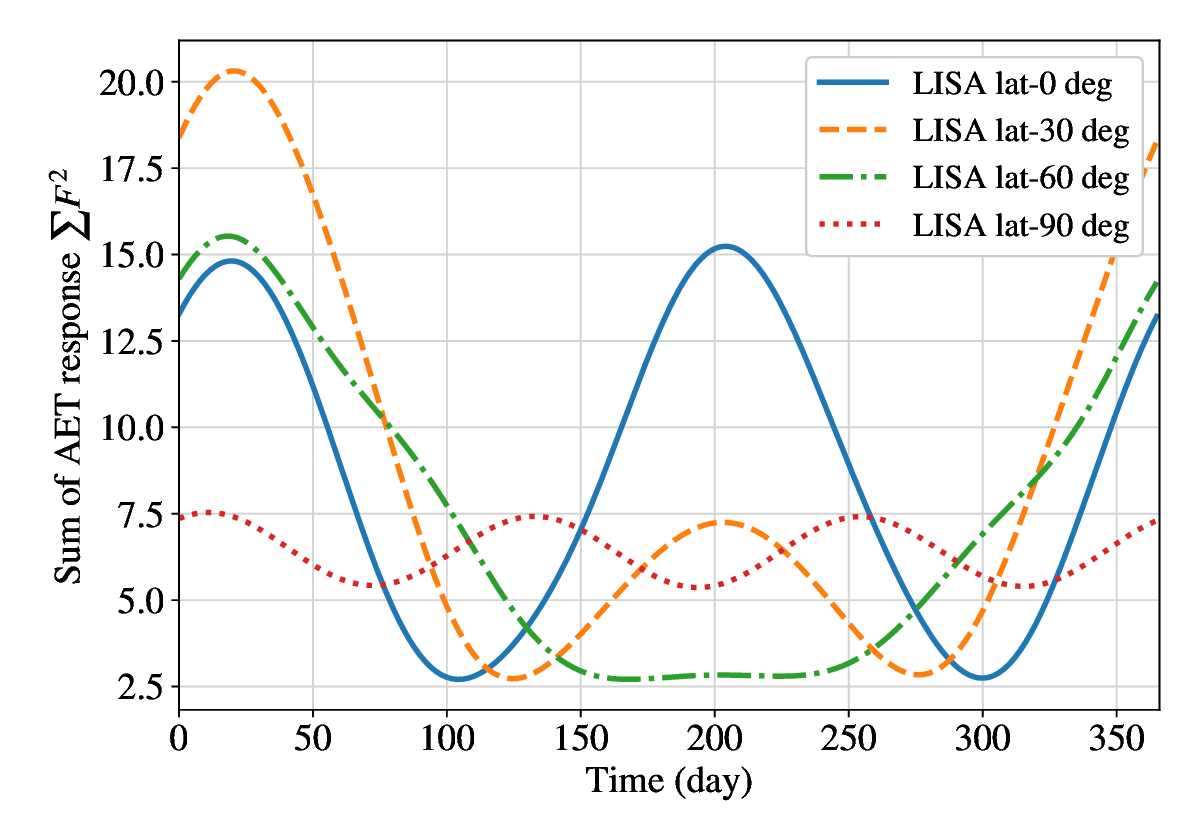}
\caption{\label{fig:response_latitude_AET} The instantaneous response of LISA X channels (left panel) and the sum of responses in LISA's A, E and T channels varying with yearly orbital motion at selected ecliptic latitudes, $\sum_{\rm AET} |F^{h}_{\rm TDI} (f=20\ {\rm mHz}, \psi = 0, \iota = 0)|^2$ (right panel) to monochromatic 20 mHz signal.}
\end{figure}

The instantaneous response function $F^h_\mathrm{X} (\lambda, \beta, \psi, \iota, f)$ of the Michelson-X channels to a specific GW source (location $\lambda$ and $\beta$, polarization $\psi$, inclination $\iota$, and frequency $f$) could be obtained by substituting \Eref{eq:y_ij} into \Eref{eq:eta} and \Eref{eq:X_measurement}. The most sensitive directions of a detector are around the normal direction of the plane formed by three S/C as shown in the left plot of Figure \ref{fig:response_latitude_AET}.
In one orbital period, the sensitive regions will change with the constellation's motion and rotation. And the response will modulate for a source. By selecting four latitudes ($0^\circ$, $30^\circ$, $60^\circ$, and $90^\circ$) on a longitude, the sum of response in the A, E and T channels at 20 mHz, $\sum_{\rm AET} F^2(f=20\ {\rm mHz}, \psi = 0, \iota = 0)$, are shown in the right panel of Figure \ref{fig:response_latitude_AET}. The responses at different latitudes are rather different and change with the orbit motion periodically. For a monochromatic GW signal in a TDI channel, the area between $y=0$ axes and a curve should be proportional to the square of SNR $\rho^2$ (since the sensitivity of T channel at 20 mHz is poor, and the A and E channels have the same noise PSD).

\begin{figure}[htb]
\includegraphics[width=0.48\textwidth]{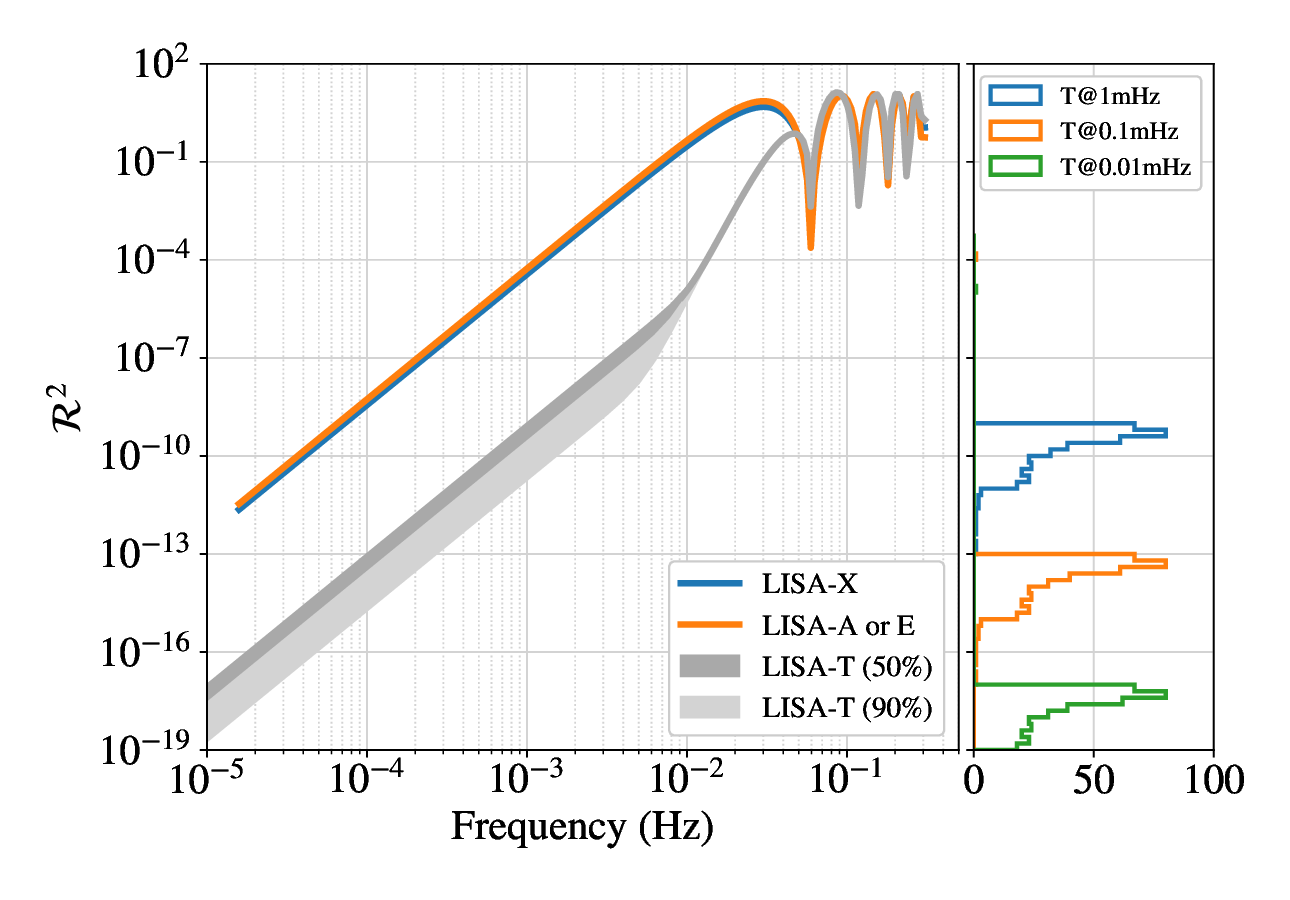}
\includegraphics[width=0.48\textwidth]{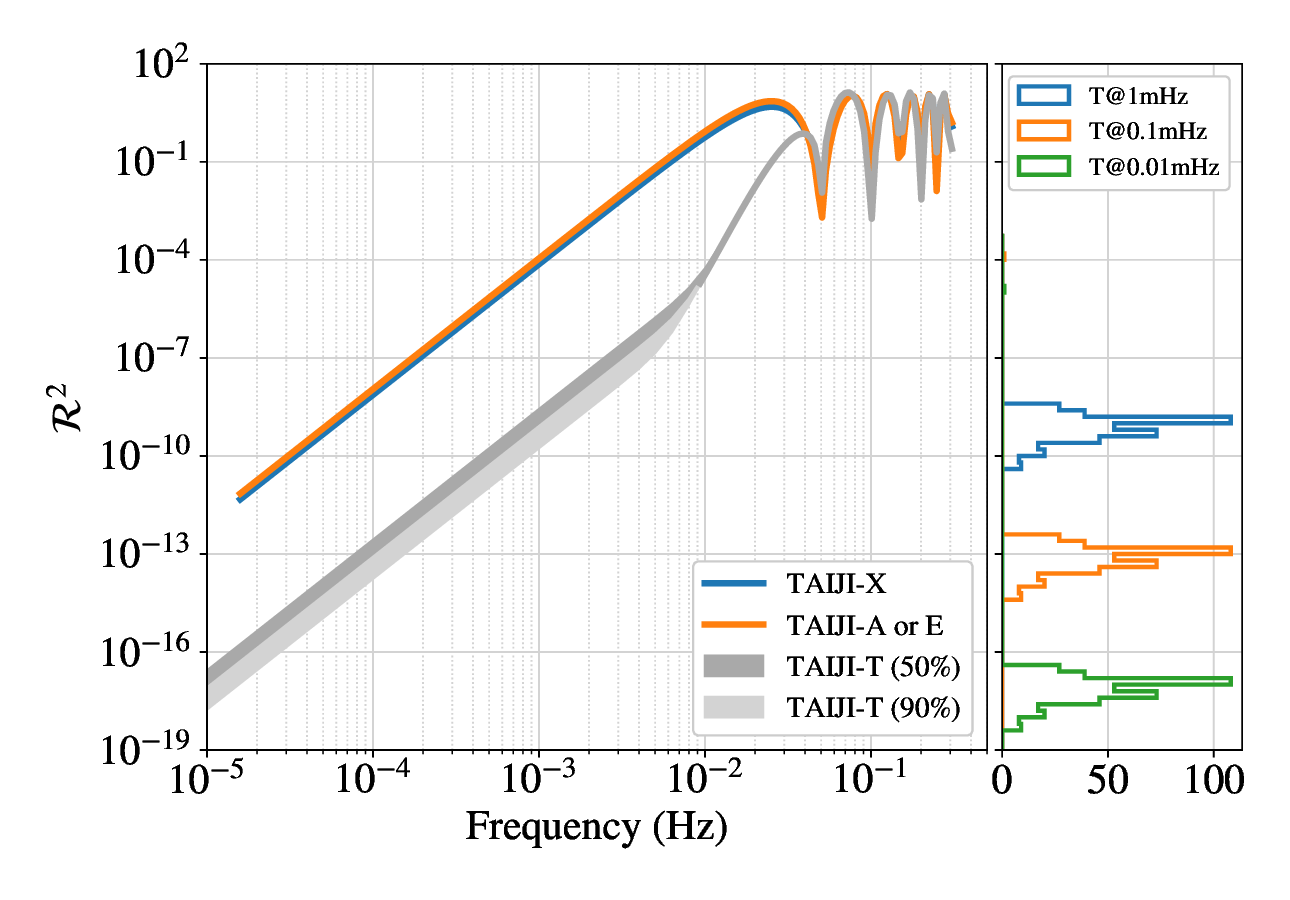}
\caption{\label{fig:response_1st} The average responses of TDI X, A, E, and T channels for LISA (left panel) and TAIJI (right panel). The T channel is sensitive to the variances of arm lengths, and the dark grey region shows the best 50\% percentile in 400 days, and the dark and light grey areas together show the best 90\% percentile. The right panel in each plot shows the histogram of the T channel's response at frequencies 0.01 mHz, 0.1 mHz and 1 mHz.}
\end{figure}

To evaluate the detectability of a TDI channel, the averaged response over sky and polarization at each frequency is calculated, 
\begin{eqnarray} \label{eq:resp_averaged}
 \mathcal{R}^2_{\rm TDI} (f) =& \frac{1}{4 \pi^2}  \int^{2 \pi}_{0} \int^{\frac{\pi}{2}}_{-\frac{\pi}{2}} \int^{\pi}_{0} |F^{h}_{ \rm TDI} (f, \iota=0)|^2 \cos \beta {\rm d} \psi {\rm d} \beta {\rm d} \lambda.
\end{eqnarray}
The average responses for LISA and TAIJI X, A, E and T are shown in Figure \ref{fig:response_1st}. The A and E curves are identical and higher than X channel by a factor of $3/2$ as can be inferred from \Eref{eq:equivalent_A} and \Eref{eq:equivalent_E}. The averaged GW responses of X, A and E could be approximated as \cite{Cornish:2018dyw},
\begin{eqnarray}
\mathcal{R}^2_{\rm X} (f) \simeq  \frac{3}{10} \frac{16 x^2 \sin^2 x}{1 + 0.6 x^2}, \\
\mathcal{R}^2_{\rm A} (f) = \mathcal{R}^2_{\rm E} (f) \simeq \frac{9}{20} \frac{16 x^2 \sin^2 x}{1 + 0.6 x^2},
\end{eqnarray}
where $x = 2 \pi f L$. Although the average response of A and E channels are equal, their response to a source could not be the same as indicated by \Eref{eq:equivalent_A} and \Eref{eq:equivalent_E}. In low-frequency approximation ($f \ll 1 / 2 \pi L $), the response of X/A/E channel should be proportional to $x^4$, and the response variation caused by the unequal arm is negligible since the arm length variations are less than 1\% as shown in Figure \ref{fig:orbit}.

\begin{figure}[htb]
\centering
\includegraphics[width=0.48\textwidth]{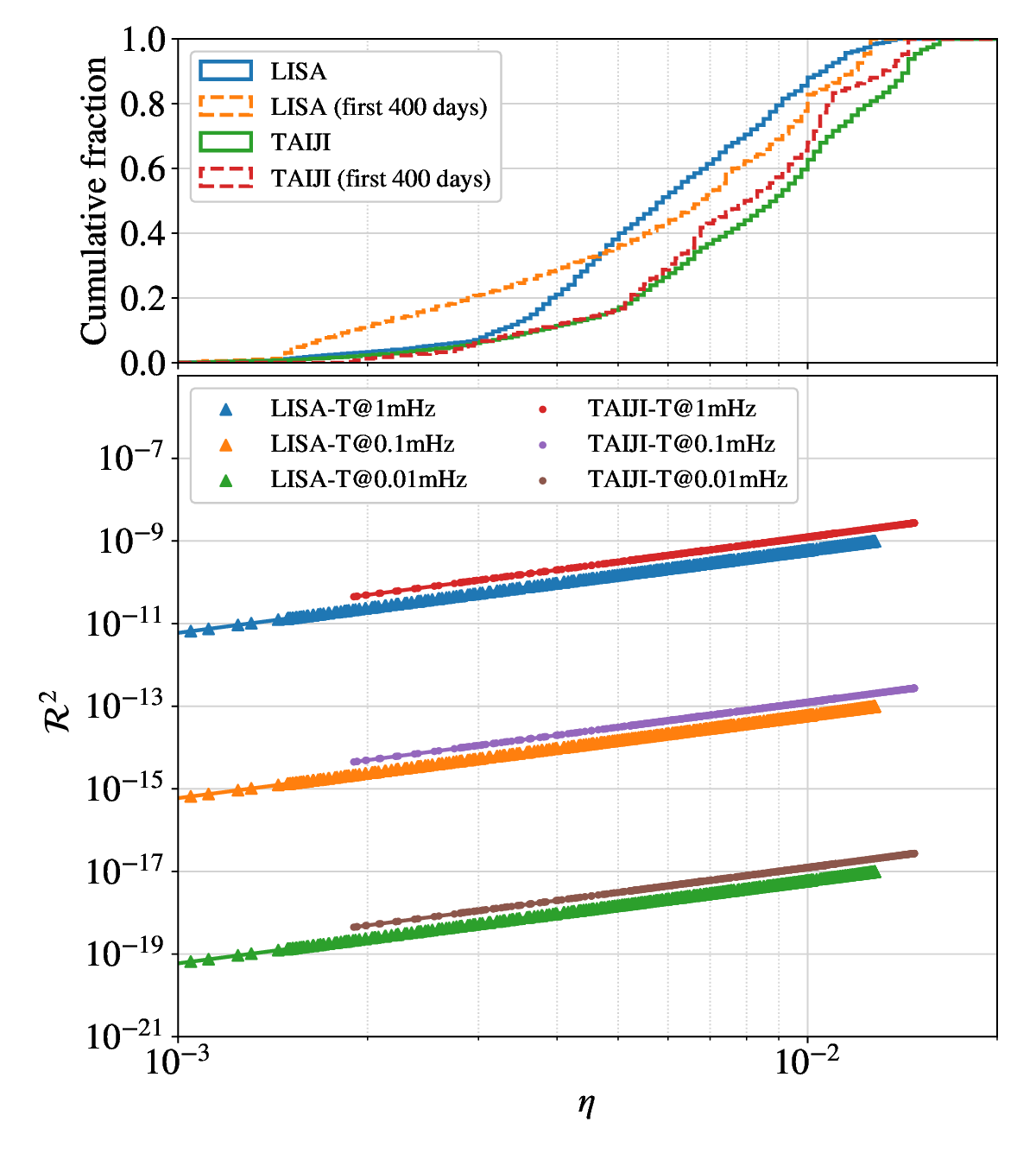} 
\caption{\label{fig:inequality_and_fit} The histogram of inequality $\eta$ in the full period and first 400 days of the LISA and TAIJI mission orbits (upper panel), and the relation between the inequality factor $\eta$ and the response of T channel at selected frequency (0.01 mHz, 0.1 mHz and 1 mHz).}
\end{figure}

The T channel is more sensitive to the unevenness of arms, the dark grey area shows the best 50\% percentile of the response in the first 400 days of the numerical orbit, and the dark with light grey area shows the best 90\% percentile. The right panel in each plot shows the histograms of response at the frequency 0.01 mHz, 0.1 mHz, and 1 mHz, respectively. And the response of T channel is proportional to the {\it inequality} factor of the arm lengths,
\begin{equation} \label{eq:inequality}
 \eta (t) \equiv \frac{\sqrt{ (L_{12} - L_{23})^2 + (L_{12} - L_{13})^2 + (L_{13} - L_{23})^2 }}{L}.
\end{equation}
The inequality factor for the full period and the first 400 days are shown in the upper panel of Figure \ref{fig:inequality_and_fit}, and the values are mainly in a range of $[0.001, 0.02 ]$. As the response of T channel versus inequality is shown in the lower panel, we can notice their linear relation in the log-log plot. The response of T channel, $\mathcal{R}^2_{\rm T}$, should be proportional to $\eta^2$ and $x^4$, and a fitting formula is obtained,
\begin{equation} \label{eq:T_response_eta_fit}
 \mathcal{R}^2_{\rm T} (f) \approx 0.8 x^4 \eta^2.
\end{equation}

This approximation would be applicable for frequencies lower than $\sim$5 mHz. The inequality of arm lengths, $\eta(t)$, could be obtained from the numerical orbit, and then the GW response of T channel could be calculated. 

\subsection{Noises in TDI Channels} \label{sec:noise}

\subsubsection{Laser frequency noise}

The laser frequency noise in Michelson-X channel could be obtained by substituting the laser noise components in \Eref{eq:eta}-\Eref{eq:s_epsilon_tau_2} into \Eref{eq:X_measurement},
\begin{equation}
 \mathrm{X}_{\rm laser} \simeq \dot{C}_{12}  \delta t_\mathrm{X},
\end{equation}
where $\dot{C}_{12}$ is the time derivative of laser frequency noise on S/C1 optical bench pointing to S/C2, and $\delta t_\mathrm{X}$ is the time difference between the two path lengths shown in Figure \ref{fig:X_U_P_D_diagram}. The PSD of residual laser noise on the Michelson-X channel is estimated by \Eref{eq:dC}. Two approaches are implemented to calculate the path mismatches in TDI channels.

\begin{figure}[htb]
\includegraphics[width=0.46\textwidth]{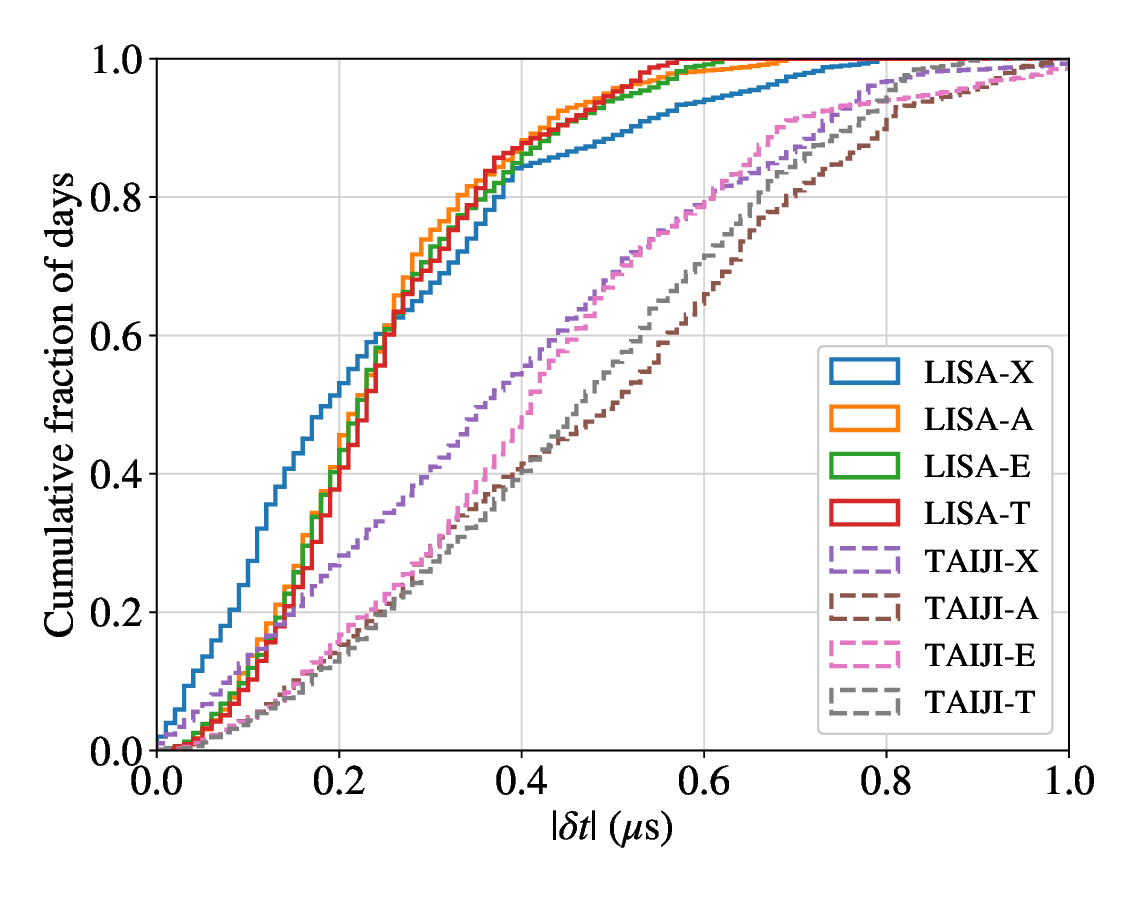}
\includegraphics[width=0.48\textwidth]{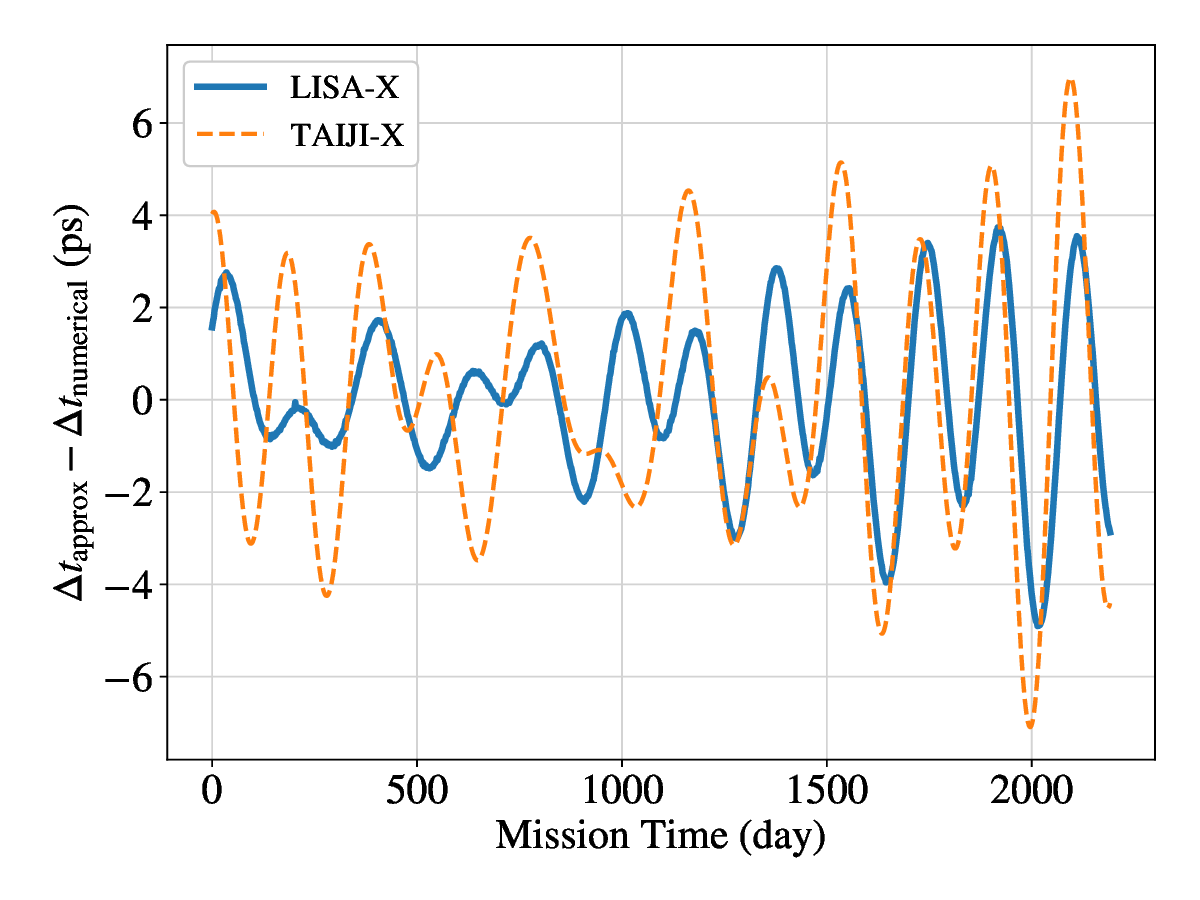}
\caption{\label{fig:TDI_dt_1st} The cumulative histogram of the time mismatches for X, A, E, and T channels interferometry paths in the 2200 days (left panel), and the time difference between the numerical method and approximated method (right panel). }
\end{figure}

The first approach is the numerical method which calculates the light propagation time along each arm in time sequential order in TDI observable. The position of a receiver S/C is determined by iterative interpolation, and the time delay from relativistic effect is also incorporated into the calculation. Our numerical method was initially developed for the TDI calculation in ASTROD-GW concept \cite{Wang:2011,Wang:2012te,Ni:2013,Wang:2014aea,Wang:2014cla}, and was applied to the LISA and TAIJI mission subsequently \cite{Wang:2012ce,Dhurandhar:2011ik,Wang:2017aqq,Wang:2020cpq}. The numerical method can provide high accuracy results especially for long interferometric arms. The cumulative histogram of time mismatches for X, A, E, and T channels are shown in the left panel of Figure \ref{fig:TDI_dt_1st}, and the values for the A, E, and T channels are inferred from \Eref{eq:AET_laser_noise}. In 2200 days, the time differences are mostly less than 0.8 $\mu$s for LISA, and the time differences are less than 1.0 $\mu$s for TAIJI mission.

The second algorithm is to approximate the light propagation time calculation into the arm length and its first derivative with respect to time as shown in \Eref{eq:dL_X_approx}. 
\begin{eqnarray} \label{eq:dL_X_approx}
\delta t_{\rm X}(t) & = [ L_{31}(t) + \mathcal{D}_{31} L_{13} (t) + \mathcal{D}_{13} \mathcal{D}_{31} L_{21} (t) + \mathcal{D}_{21} \mathcal{D}_{13} \mathcal{D}_{31} L_{12} (t) ] \nonumber \\ & \quad  - [ L_{21}(t) + \mathcal{D}_{21} L_{12} (t) + \mathcal{D}_{12} \mathcal{D}_{21} L_{31} (t) + \mathcal{D}_{31} \mathcal{D}_{12} \mathcal{D}_{21} L_{13} (t) ] \nonumber \\
& \simeq  -L_{31} \dot{L}_{13} - ( L_{13} + L_{31} ) \dot{L}_{21} - (L_{21} + L_{13} + L_{31} ) \dot{L}_{12} 
 +L _{21} \dot{L}_{12} \nonumber \\ 
 & + ( L_{12} + L_{21} ) \dot{L}_{31} + (L_{31} + L_{12} + L_{21} ) \dot{L}_{13} \nonumber\\
& \simeq  4 L_{12} \dot{L}_{13} - 4 L_{13} \dot{L}_{12}.
\end{eqnarray}
As we verified, this approach could achieve a qualified precision for the first-generation TDI calculation for LISA and TAIJI since the $3L$ ($\simeq 25$ s for LISA and $\simeq 30$ s for TAIJI) time delay is short compared to the relative motions between S/C. As the right plot shown in Figure \ref{fig:TDI_dt_1st}, the difference of results from two algorithms is within 8 ps for the X channel. A caveat is that the precision of approximation could be declined with the increase of arm length and/or total propagation time, and it has been reflected in the plot comparing the longer arm length of TAIJI ($3\times 10^6$ km) and relatively short arm LISA ($2.5 \times 10^6$ km). Also for the second-generation TDI configurations, with the increase of propagation time, the accuracy of the approximate algorithm could further degenerate.

\begin{figure*}[htb]
\includegraphics[width=0.48\textwidth]{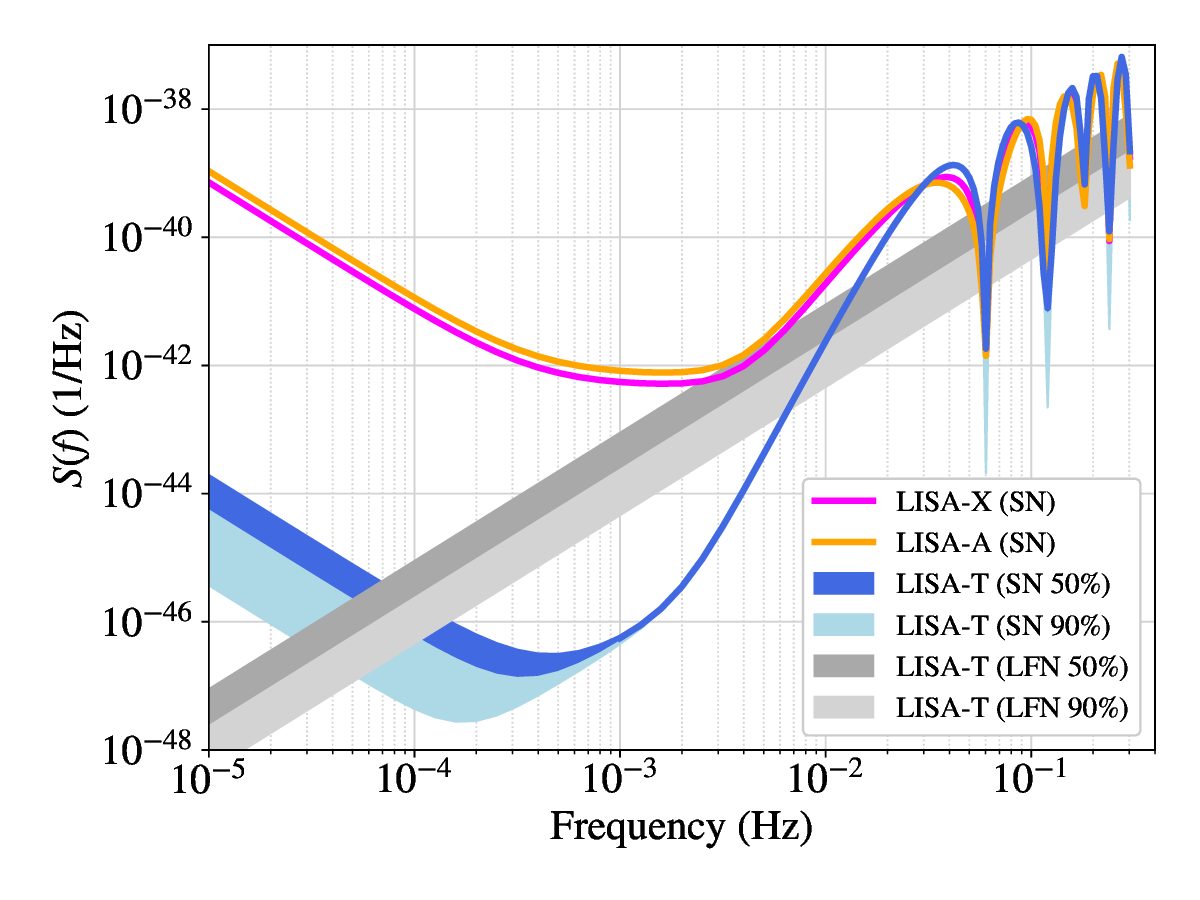}
\includegraphics[width=0.48\textwidth]{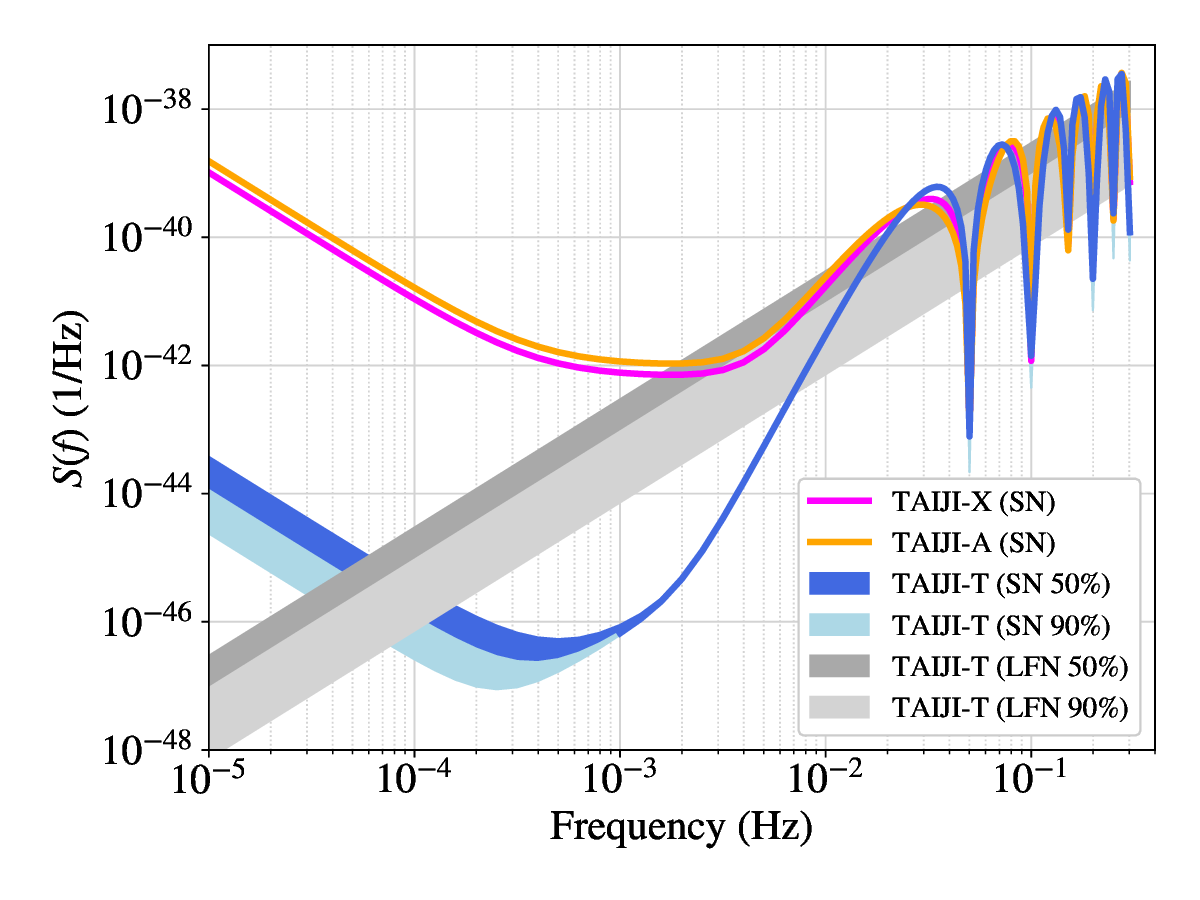}
\caption{\label{fig:Sn_laser_secondary_noise} The residual laser frequency noise (LFN) and secondary noise (SN, acceleration noise+optical metrology noise) for LISA (left panel) and TAIJI (right panel) in the first 400 days. The dark grey shows the highest laser noise in 50\% of the first 400 days, and dark grey together with light grey show the laser noise 90\% percentile in T channel. The dark blue shows the highest secondary noise in 50\% of the first 400 days in T channel, and the dark blue together with light blue show the secondary noise in 90\%. }
\end{figure*}

After the time mismatches are obtained, the residual laser noise in each TDI channel could be evaluated. Considering the mismatches are in the same levels, the PSD of the residual laser noise in X, A, E and T channels should also be comparable. And T channel is selected to represent the laser noise as shown in Figure \ref{fig:Sn_laser_secondary_noise}. The dark grey area shows the highest noise level in 50\% of the first 400 days, and the dark and light grey area shows the highest noise level in 90\% percentile.

\subsubsection{Secondary noise} \label{secsub:TDI_noise}

The acceleration noise and optical path noise are treated as secondary noise. And the noises could be evaluated by substituting the corresponding terms in \Eref{eq:eta}-\Eref{eq:s_epsilon_tau_2} into \Eref{eq:X_measurement}. The noise PSD of the optimal channels A, E and T could be obtained from the combinations of \Eref{eq:optimalTDI}. Based on the investigation, the noise in X/A/E channel is insensitive to the arm inequality. However, the noise PSD of T channel diverges from the equal-arm and varies with the inequality of arm lengths for the low frequencies. Their PSD functions for a static unequal arm ($L_{ij}=L_{ji}$ and $L_{12} \neq L_{13} \neq L_{23} \neq L_{12}$) configuration could be formulated as follows,
\noindent
\begin{equation} \label{eq:Sn_XAET}
\eqalign{
\fl S_\mathrm{X}(f) =  8 S_{\mathrm{op}} \left( \sin^2  x_{12}  +  \sin^2  x_{13}  \right) + 32 S_\mathrm{acc} \left( 1 - \cos^{2}{x_{12} } \cos^{2}{x_{13} }  \right),  \cr
\fl S_\mathrm{A}(f) =   4 S_\mathrm{op} \left[ \sin^{2}{x_{12} } + 2 \sin^{2}{x_{13} } + \sin^{2}{x_{23} } + 2 \sin{x_{12} } \sin{x_{23} } \cos{x_{13} } \cos{ (x_{12} - x_{23} ) } \right] \cr
\fl \qquad  \qquad +  16 S_\mathrm{acc} \left[ 2 \sin{x_{12} } \sin{x_{23} } \cos{x_{13} } \cos{ ( x_{12} - x_{23} ) } - \cos^{2}{x_{12} } \cos^{2}{x_{13} } - \cos^{2}{x_{13} } \cos^{2}{x_{23} } + 2 \right],  \cr
\fl S_\mathrm{E}(f) =  \frac{4 S_\mathrm{op} }{3}  \left[ 5 \sin^{2}{x_{12} } + 2 \sin^{2}{x_{13} }  + 5 \sin^{2}{x_{23} } + 4 \sin{x_{13} } \sin{x_{23} } \cos{x_{12} } \cos{( x_{13} - x_{23} ) }  \right. \\
 \left.   - 2 \sin{x_{12} } \sin{x_{23} } \cos{x_{13} } \cos{( x_{12} - x_{23} ) } + 4 \sin{x_{12} } \sin{x_{13} } \cos{x_{23} } \cos{ ( x_{12} - x_{13} ) } \right] \\
\fl \qquad  \qquad  +  \frac{16 S_\mathrm{acc}}{3} \left[6  - \cos^{2}{x_{12} } \cos^{2}{x_{13} } - 4 \cos^{2}{x_{12} } \cos^{2}{x_{23} } - \cos^{2}{x_{13} } \cos^{2}{x_{23} } \right. \\
  \left.  + 4 \sin{x_{12} } \sin{x_{13} } \cos{x_{23} } \cos{( x_{12} - x_{13} ) }  - 2 \sin{x_{12} } \sin{x_{23} } \cos{x_{13} } \cos{( x_{12} - x_{23} )} \right. \\
  \left.  + 4 \sin{x_{13} } \sin{x_{23} } \cos{x_{12} } \cos{( x_{13} - x_{23} ) }  
  \right], \\
\fl S_\mathrm{T}(f) =  \frac{16 S_\mathrm{op}  }{3} \left[ \sin^{2}{x_{12} } + \sin^{2}{x_{13} } + \sin^{2}{x_{23} } - \sin{x_{12} } \sin{x_{13} } \cos{x_{23} } \cos{( x_{12} - x_{13} ) }  \right. \\ \left.  - \sin{x_{12} } \sin{x_{23} } \cos{x_{13} } \cos{( x_{12} - x_{23} ) }  - \sin{x_{13} } \sin{x_{23} } \cos{x_{12} } \cos{( x_{13} - x_{23} ) } \right] \\
\fl \qquad  \qquad   + \frac{32 S_\mathrm{acc}}{3} \left[ 3 - \cos^{2}{x_{12} } \cos^{2}{x_{13} } - \cos^{2}{x_{12} } \cos^{2}{x_{23} } - \cos^{2}{x_{13} } \cos^{2}{x_{23} } \right. \\
 \left.  - 2 \sin{x_{12} } \sin{x_{13} } \cos{x_{23} } \cos{ ( x_{12} - x_{13} ) }   - 2 \sin{x_{12} } \sin{x_{23} } \cos{x_{13} } \cos{( x_{12} - x_{23} ) }   \right. \\
 \left. - 2 \sin{x_{13} } \sin{x_{23} } \cos{x_{12} } \cos{ ( x_{13} - x_{23} )} \right],  
}
\end{equation}
where $x_{ij}=2\pi f L_{ij} (i, j = 1,2,3)$. By applying the noise budgets, their PSD curves for LISA and TAIJI are shown in Figure \ref{fig:Sn_laser_secondary_noise}. The dark blue area shows the highest 50\% percentile of noise PSD of T channel in the first 400 days, and the dark blue together with the light blue areas show the noise level in 90\% percentile.

\subsection{Sensitivities of Michelson configuration} \label{sec:sensitivity}

The average sensitivity of a TDI channel could be obtained by weighting its noise PSD with the average response, $S_{\rm avg} = {S_{\rm n}}/{\mathcal{R}^2}$. To understand the impacts of laser noise and secondary noise, we examine the sensitivities with secondary noise only first, and the sensitivity from secondary noise together with laser frequency noise thereafter.

\begin{figure*}[htb]
\includegraphics[width=0.48\textwidth]{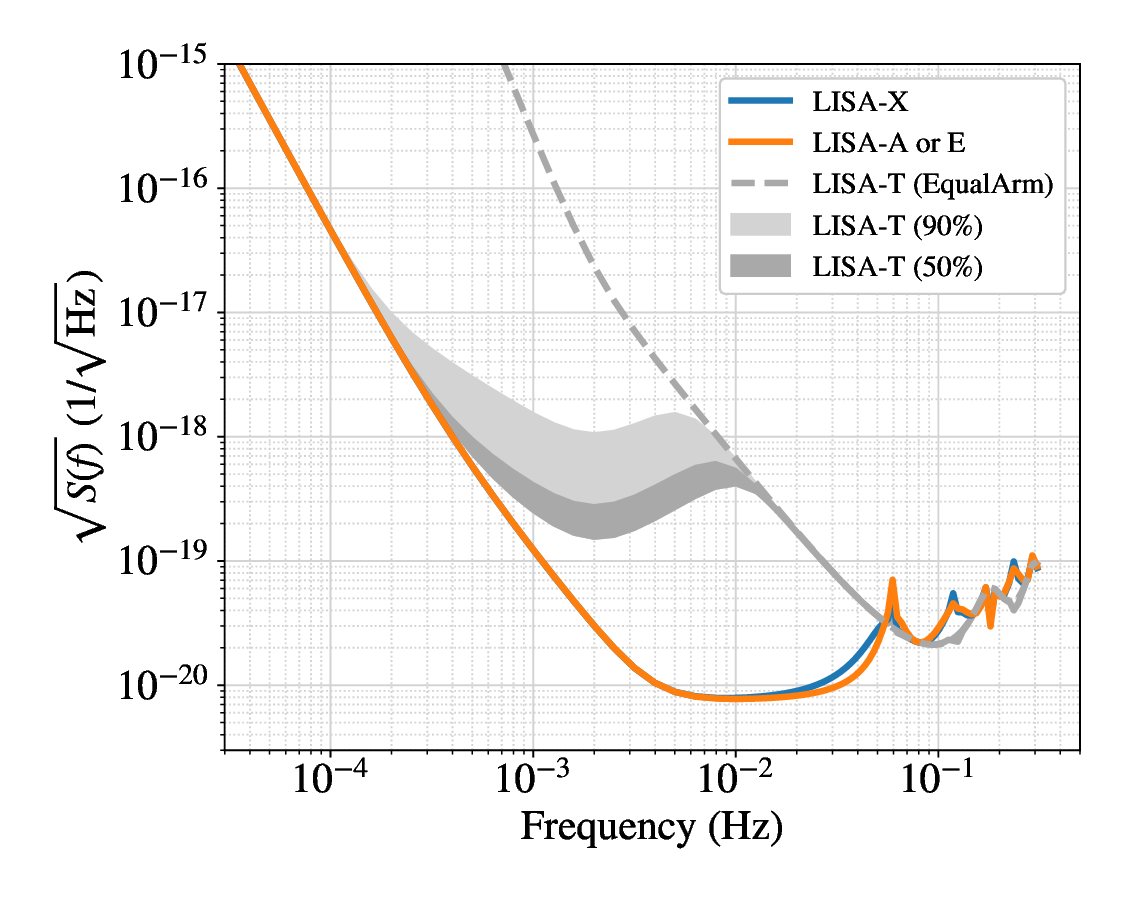}
\includegraphics[width=0.48\textwidth]{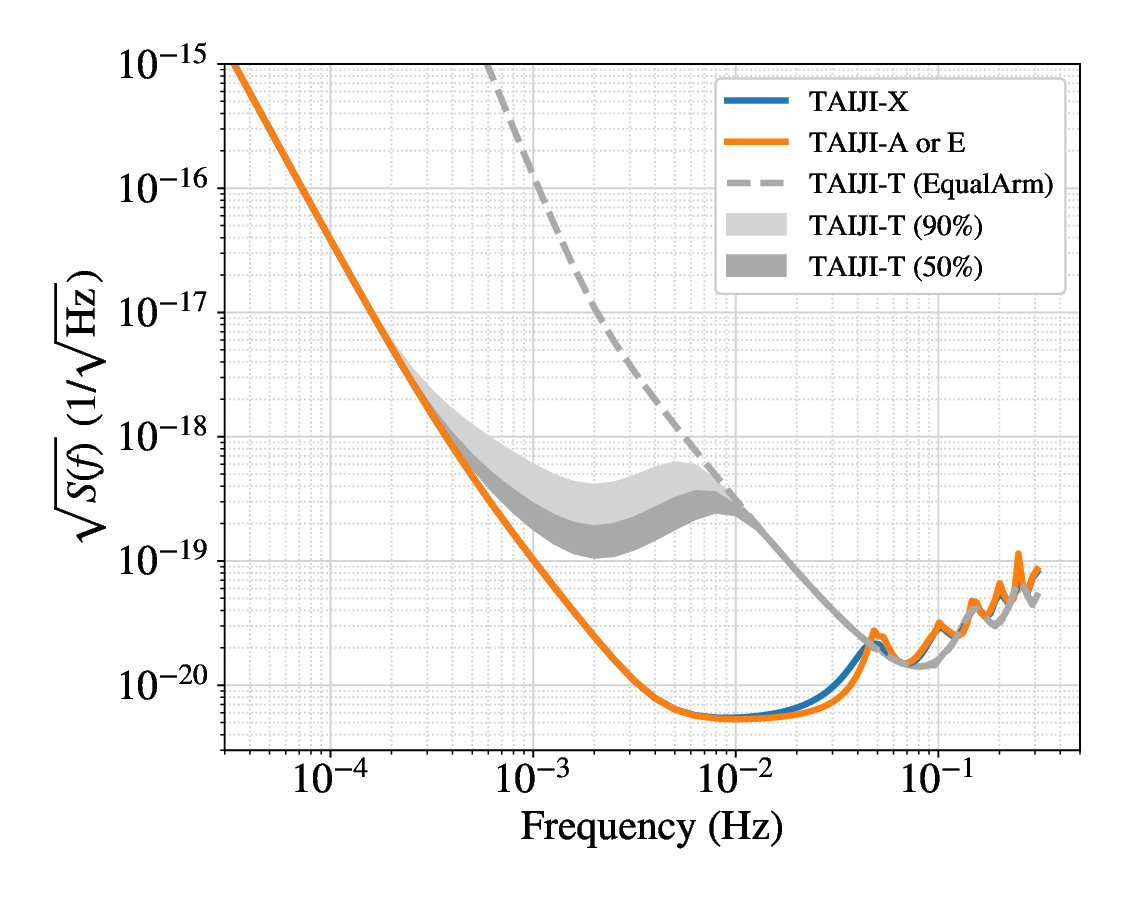}
\caption{\label{fig:Sensitivity_secondary_noise} The average sensitivities of TDI channels X, A, E, and T for LISA (left) and TAIJI (right) considering the secondary noises. The dark grey area shows the best sensitivity of T channel in the 50\% percentile of the first 400 days, and the dark grey together with the light grey areas show the best sensitivity in 90\% percentile. The sensitivities of T channels from equal-arm configuration are shown by curves LISA-T (EqualArm) and TAIJI-T (EqualArm) \cite{Prince:2002hp,Vallisneri:2007xa}.}
\end{figure*}

When only secondary noise is considered, the average sensitivities of X, A, E, and T channels are shown in Figure \ref{fig:Sensitivity_secondary_noise}. The sensitivities of TAIJI are slightly better than the LISA's at frequencies around 10 mHz because of lower optical path noise and longer arm length. The A/E channel is slightly better than the X channel in the frequency band of [10, 50] mHz. The T channel diverges from other channels in the frequency range of [0.2, 50] mHz.
Compared to the sensitivity of T channel in the equal-arm case (grey dashed line) \cite{Prince:2002hp,Vallisneri:2007xa}, the sensitivity of T from an unequal-arm configuration is enhanced for frequencies lower than 10 mHz. And its sensitivity becomes equivalent to other channels for frequencies lower than $\sim$0.2 mHz. The dark grey area shows the best sensitivity of T channel in the 50\% percentile of the first 400 days, and the dark and light grey areas show the best 90\% percentile. 

\begin{figure*}[htb]
\includegraphics[width=0.48\textwidth]{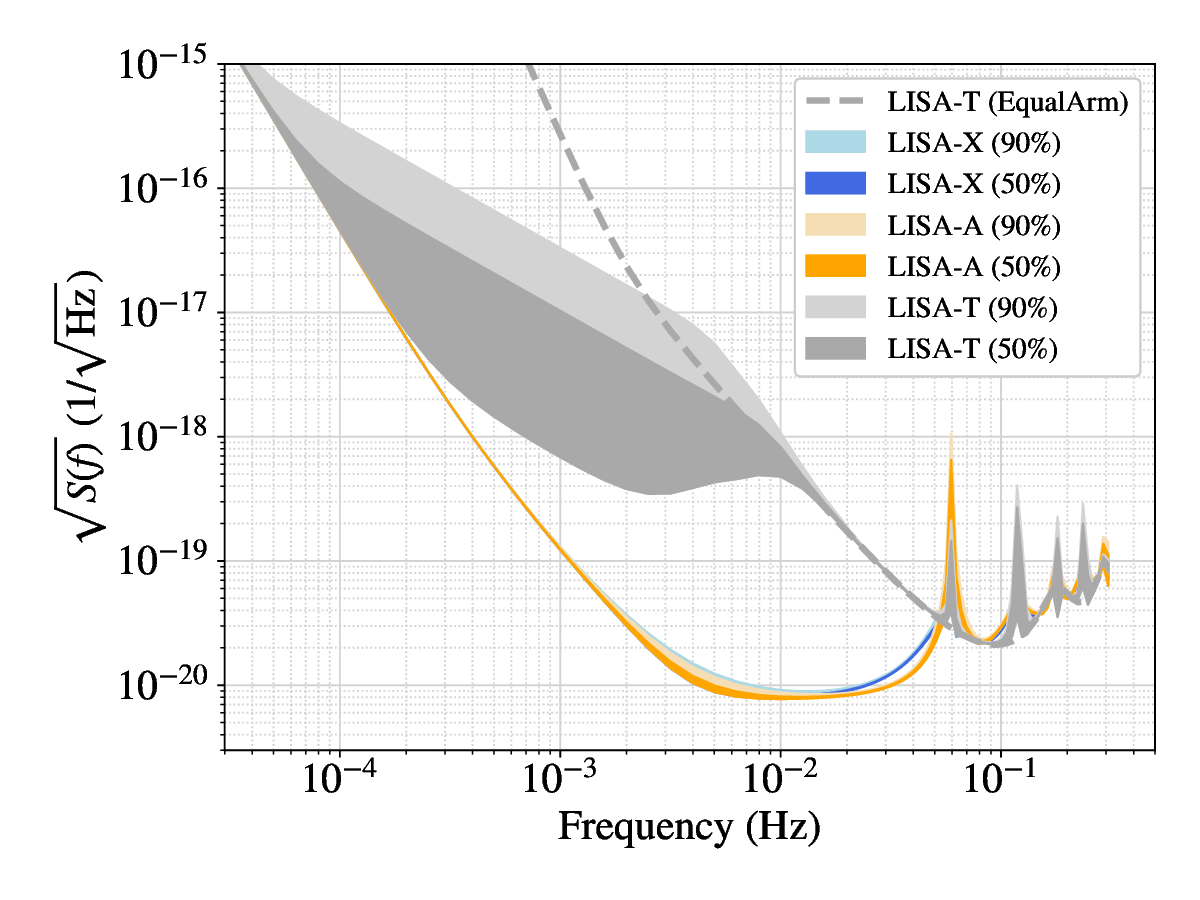}
\includegraphics[width=0.48\textwidth]{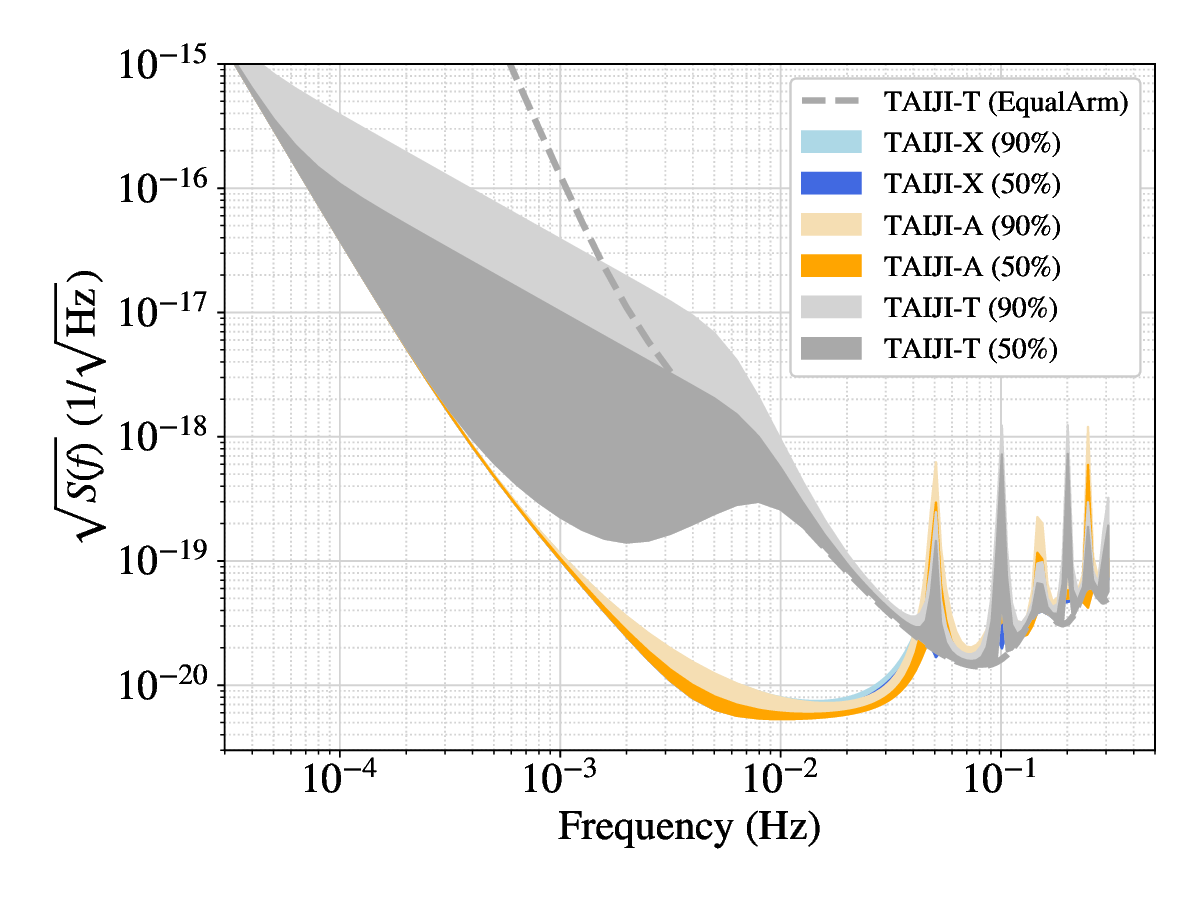}
\caption{\label{fig:Sensitivity_with_laser_noise} The average sensitivities of TDI X, A, E, and T channels by considering the secondary noises and residual laser noise. The dark color areas show the best sensitivity in the 50\% percentile of the first 400 days for X, A, and T channels, and the dark color together with the light color areas show the best sensitivity in 90\% percentile. The sensitivities of T channels for equal-arm configuration are shown by dashed curves \cite{Prince:2002hp,Vallisneri:2007xa}. }
\end{figure*}

If the residual laser noise is included together with secondary noise, the average sensitivities are shown in Figure \ref{fig:Sensitivity_with_laser_noise}. The dark and light colors show the 50\% and 90\% percentiles of sensitivities in the first 400 days. The sensitivity of the T channel deteriorates significantly because of the laser noise. At the high frequency band, all four channels are subject to the laser frequency noise at their characteristic frequencies ($f = \frac{n}{2L}$ where $n=1, 2, 3 \ldots$). Since the LISA and TAIJI have different arm lengths, and their combined sensitivities could complement each other. The joint two missions will improve the observation sensitivities by a factor of more than $\sqrt{2}$ compared to single LISA \cite{Wang:2020a,Wang:2021mou}.

\section{Optimal channels from other first-generation TDI configurations} \label{sec:other_TDI_configurations}

In this section, following the procedures in the previous section, we investigate the optimal channels from other first-generation TDI configurations in the unequal-arm scenario.

\subsection{Sagnac channels}

The measurement expression of the $\alpha$ channel from the Sagnac configuration is \cite{1999ApJ...527..814A,2000PhRvD..62d2002E},
\begin{equation} \label{eq:alpha_expression}
\alpha = (\eta_{31} + \mathcal{D}_{31} \eta_{23} + \mathcal{D}_{31} \mathcal{D}_{23} \eta_{12} ) - ( \eta_{21} +  \mathcal{D}_{21} \eta_{32} + \mathcal{D}_{21} \mathcal{D}_{32} \eta_{13} ).
\end{equation}
By substituting the laser noise terms in $\eta_{ij}$ into \Eref{eq:alpha_expression}, the residual laser term will be
\begin{equation}
 \alpha_\mathrm{laser} \simeq \dot{C}_{12} \delta t_{\alpha}
\end{equation}
where the $\delta t_{\alpha}$ is the mismatched time in the $\alpha$ channel.
Due to the Sagnac effect, for a LISA-like orbit with $60^\circ$ inclination, the time difference in a Saganc channel will be
\begin{eqnarray}
\delta t_{\alpha} = & \frac{4}{c} \vec{\omega} \cdot \vec{\mathcal{A}}  \simeq  \frac{4}{c} \mathbf{\omega} \mathcal{A} \cos 60^\circ =  \frac{2}{c} \mathbf{\omega}  \mathcal{A}, 
\end{eqnarray}
where $\mathcal{A}$ is the area of the constellation triangle, and $\omega$ is the angular velocity of rotation which is $\sim 2\pi/$yr. Therefore, the time mismatch for LISA is $\sim$12 $\mu$s as shown in Figure \ref{fig:TDI_dt_1st}, and the time mismatch for TAIJI is $\sim$17.2 $\mu$s. The absolute mismatches in Sagnac channels will be one order higher than the Michelson channels, the relative mismatches, $\max(|\delta t_{\alpha}|) - \min(| \delta t_{\alpha} |)$, is less than $\sim$0.6 $\mu$s.

\begin{figure}[htb]
\includegraphics[width=0.46\textwidth]{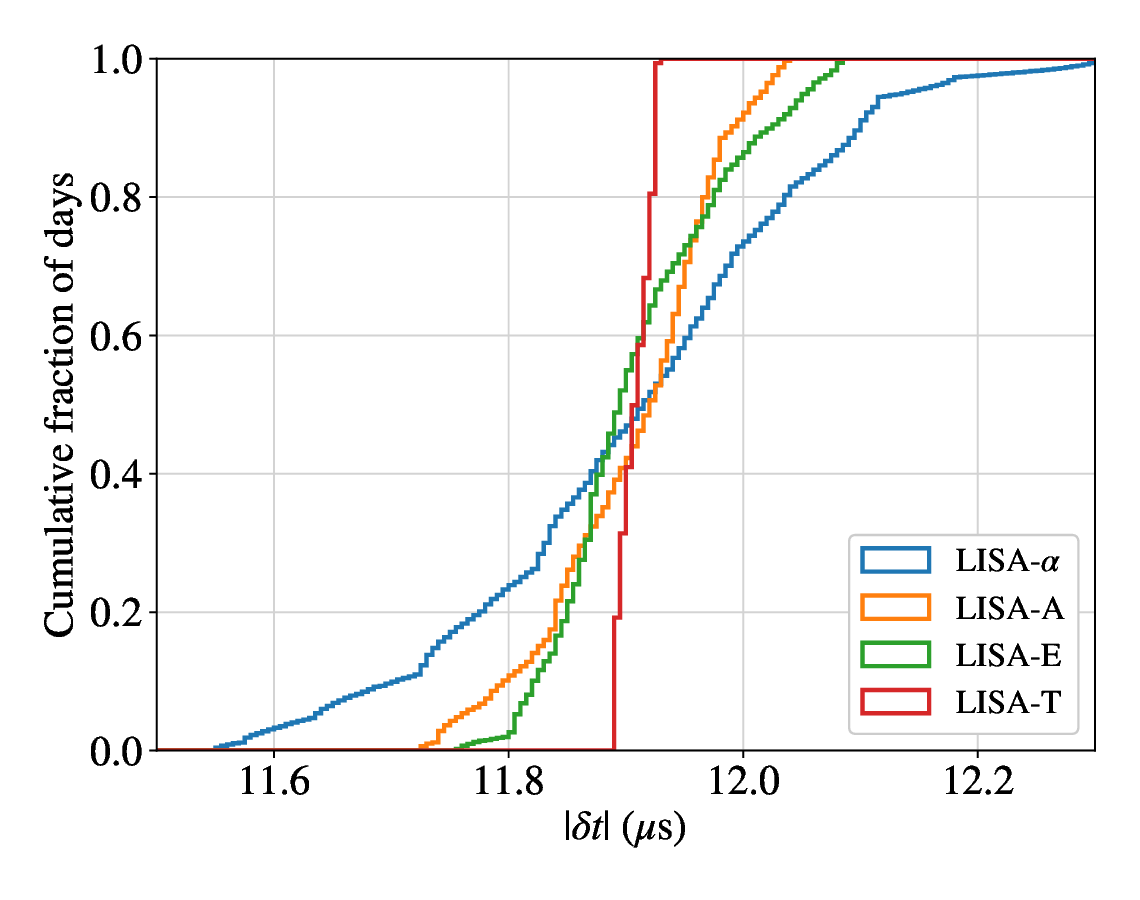}
\includegraphics[width=0.48\textwidth]{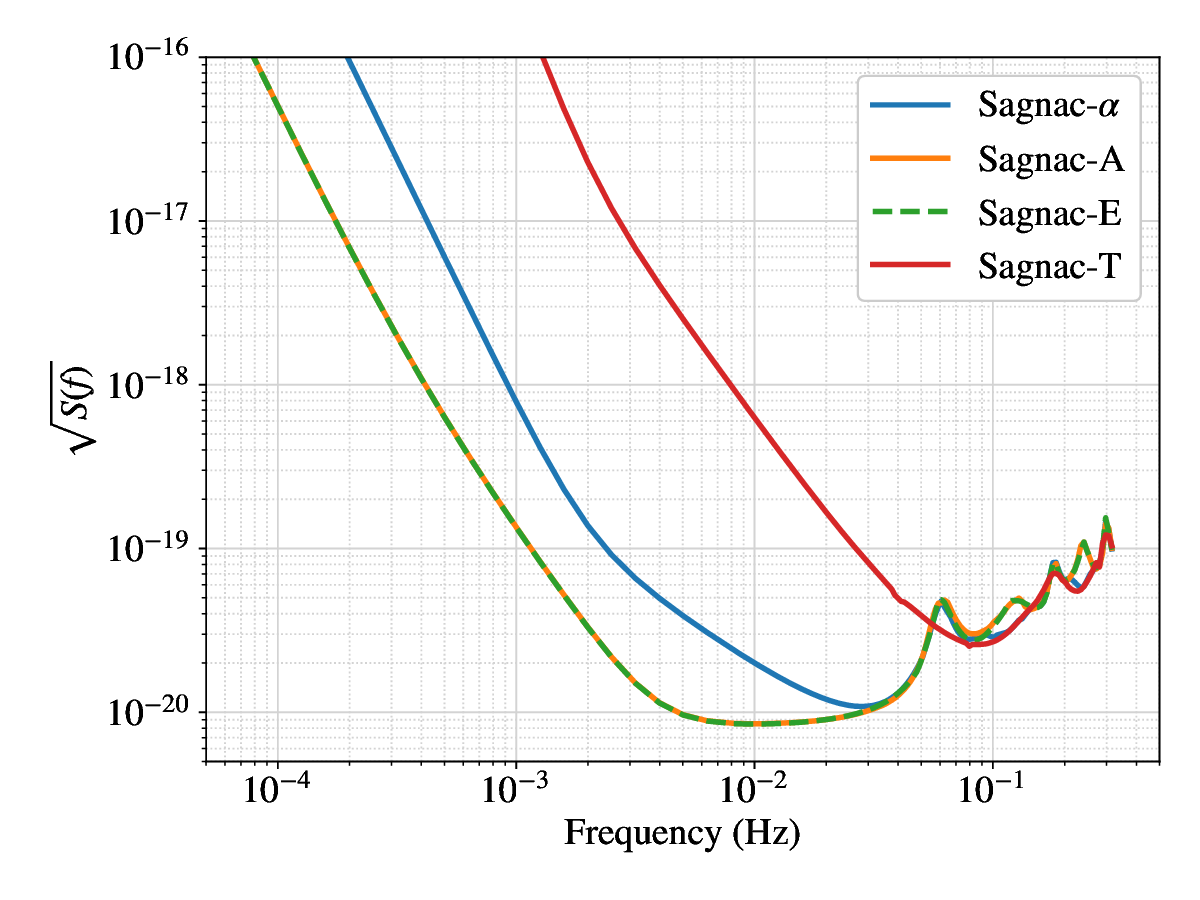}
\caption{\label{fig:Sagnac_curves} The histogram of time difference (left panel) and median sensitivities (right panel) of Sagnac $\alpha$ and the corresponding optimal channels for LISA mission. }
\end{figure}

The PSD of secondary noise in the $\alpha$ channel for unequal-arm case is described by \Eref{eq:Sn_alpha}, and the PSDs of A, E, and T channels are described by \Eref{eq:Sn_A_E_Sagnac} and \Eref{eq:Sn_T_Sagnac}. Compared to the results in equal-arm case, the $\sim$1\% inequality is insignificant for the noise PSDs and sensitivities. The PSD for A/E channel is expressed without distinguishing the three arms. The average sensitivities of the corresponding channels are shown in the right panel of Figure \ref{fig:Sagnac_curves}. 

{\setlength{\mathindent}{0cm}
\begin{eqnarray}
S_{\alpha}(f)  = &   6 S_{\mathrm{op}} + 4 S_{\mathrm{acc}} \left[ \sin x_{23}  \sin (x_{12} + x_{13}) - \sin x_{12}  \sin x_{13}  \cos  x_{23} \nonumber \right.
 \\ & \left. - 3 \cos  x_{12}  \cos  x_{13}  \cos  x_{23}  + 3 \right], \label{eq:Sn_alpha} \\
S_\mathrm{A,Sagnac}  \simeq & S_\mathrm{E,Sagnac} \simeq  4 S_\mathrm{op} \left( \sin^{2} x - \cos x + 1 \right) + 4  S_\mathrm{acc} \left( 4 -  3 \cos x - \cos{ 3 x } \right),  \label{eq:Sn_A_E_Sagnac} \\
S_{\mathrm{T,Sagnac}} = & \frac{2 S_{\mathrm{op}} }{3} \left[ 4\cos x_{12} + 4 \cos x_{13} + 4 \cos x_{23} + 2\cos(x_{12} + x_{13}) + 2 \cos(x_{12} + x_{23}) \right. \nonumber \\
 & \left. \qquad  + 2 \cos(x_{13} + x_{23}) + 9 \right] \nonumber \\
  & + \frac{4 S_{\mathrm{acc}}}{3} \left[ \sin x_{12} \sin(x_{13} + x_{23}) + \sin x_{13} \sin x_{23} \cos x_{12} - 9 \cos x_{12} \cos x_{13} \cos x_{23}  \right. \nonumber \\
 & \left. \qquad + 2 \cos x_{12} + 2 \cos x_{13} + 2 \cos x_{23} - 2 \cos (x_{12} - x_{13}) \right. \nonumber \\
 & \left. \qquad - 2 \cos(x_{12} - x_{23}) - 2 \cos(x_{13} - x_{23}) + 9 \right]. \label{eq:Sn_T_Sagnac}
\end{eqnarray} }

\subsection{Relay channels}

\begin{figure}[htb]
\includegraphics[width=0.46\textwidth]{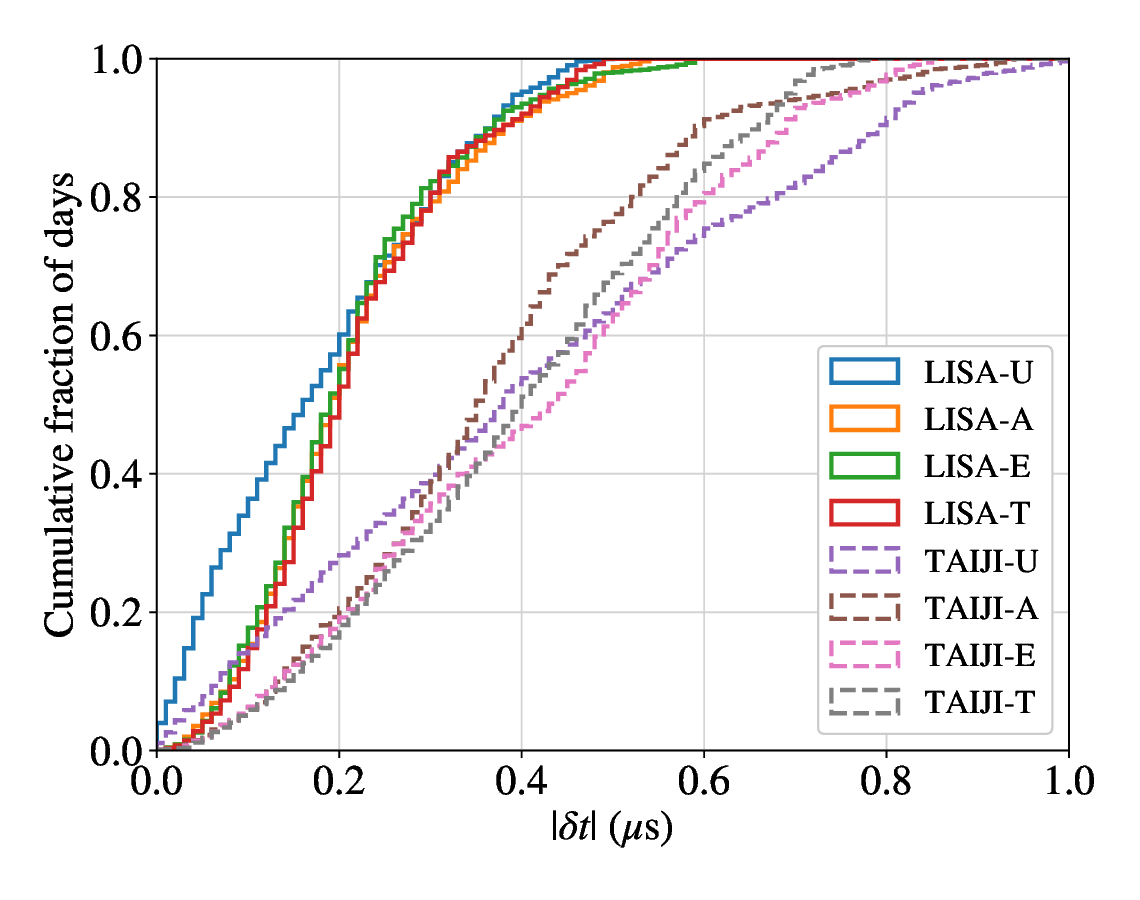}
\includegraphics[width=0.48\textwidth]{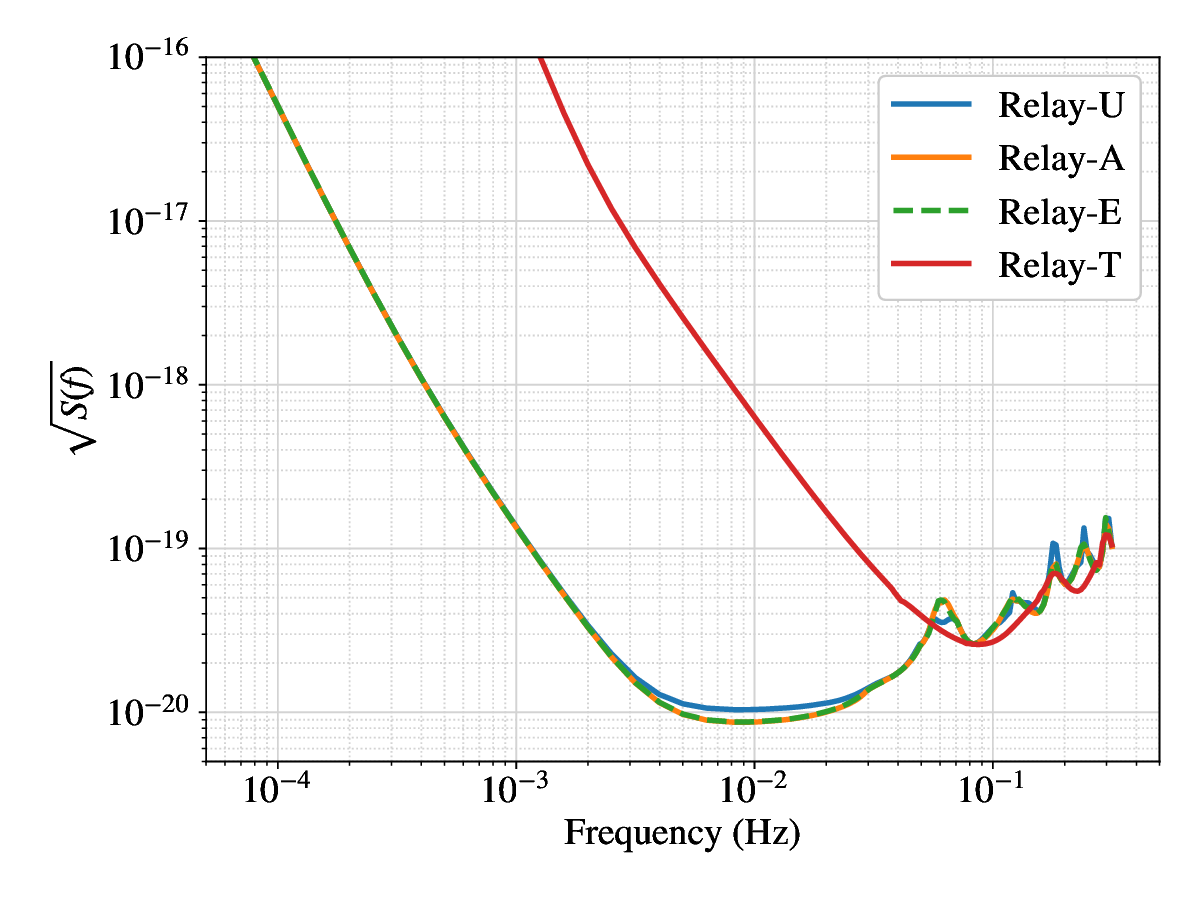}
\caption{\label{fig:Relay_curves} The histogram of time mismatches (left panel) and average sensitivities (right panel) of Relay-U and the corresponding optimal channels for LISA.}
\end{figure}

The expression of U observable from Relay (U, V, W) configuration is \cite{1999ApJ...527..814A,2000PhRvD..62d2002E},
\begin{eqnarray}
U =& (\eta_{23} + \mathcal{D}_{23} \eta_{32} + \mathcal{D}_{32}  \mathcal{D}_{23}  \eta_{13} + \mathcal{D}_{13} \mathcal{D}_{23}  \mathcal{D}_{32}  \eta_{21} ) \nonumber \\
         & - ( \eta_{13} + \mathcal{D}_{13} \eta_{21} + \mathcal{D}_{21}  \mathcal{D}_{13}  \eta_{32} + \mathcal{D}_{32}  \mathcal{D}_{21}  \mathcal{D}_{13} \eta_{23}  ).
\end{eqnarray}
The diagram of Relay-U is shown by the second plot of Figure \ref{fig:X_U_P_D_diagram}, and the residual laser noise depends on the mismatch $\delta t_\mathrm{D}$ at laser noise $C_{23}$ which is noise of laser source on the optical bench of S/C2 pointing to S/C3. By implementing the numerical algorithm, the path mismatch is obtained and the histograms are shown in the left plot of Figure \ref{fig:Relay_curves}.

For the optimal channels from Relay configuration, the A and E channels would be not real observables because of the complex coefficients in eigenvectors \cite{Vallisneri:2007xa}. However, we still implement the optimal channels for Relay configuration to compare with others and keep the integrity of all first-generation TDI. And their path mismatches relating to the residual laser noise are shown in left plots of Figure \ref{fig:Relay_curves}. 

The PSDs of Relay-U and optimal channels yielded by the acceleration noise and the optical path noise could be expressed as
{\setlength{\mathindent}{0cm}
\begin{eqnarray}
S_{\mathrm{U}} (f) = & 4 S_{\mathrm{op}} \left[ 2 \sin^{2} x_{23} - \cos x_{23} \cos(x_{12} + x_{13}) + 1 \right] \nonumber  \\ 
& + 8 S_{\mathrm{acc}} \left[ \sin^{2} x_{23} - 2 \cos x_{23} \cos(x_{12} + x_{13}) + 2 \right], \label{eq:Sn_U}  \\
S_\mathrm{A,Relay} \simeq & S_\mathrm{E,Relay} \simeq 
S_\mathrm{op} \left( 10 - 3 \cos{x } - 6 \cos{2 x} - \cos{ 3 x} \right)  \nonumber \\
& + 2 S_\mathrm{acc} \left( 13 - 4 \cos{x } - 4 \cos{2 x} - 4 \cos{3 x} - \cos{4 x}  \right), \\
S_\mathrm{T,Relay} \simeq & \frac{4 S_\mathrm{op}}{3} \left[ - \sin^{2} x_{12} \cos x_{13} + 2 \sin^{2} x_{12} + 3 \sin x_{12} \sin x_{13} \cos x_{12} - \sin x_{12} \sin x_{13} \right. \nonumber \\
 & \left. \qquad + 3 \sin x_{12} \sin x_{23} \cos x_{23} - \sin x_{12} \sin x_{23} + \sin x_{12} \sin{\left(x_{13} + x_{23} \right)} \right. \nonumber \\
 & \left. \qquad - \sin x_{12} \sin{\left(x_{12} + x_{13} - x_{23} \right)} - \sin^{2} x_{13} \cos x_{23} + 2 \sin^{2} x_{13} + \sin x_{13} \sin x_{23} \cos x_{12} \right. \nonumber \\
 & \left. \qquad + 3 \sin x_{13} \sin x_{23} \cos x_{13} - \sin x_{13} \sin x_{23} - \sin x_{13} \sin{\left(- x_{12} + x_{13} + x_{23} \right)} \right. \nonumber \\
 & \left. \qquad - \sin^{2} x_{23} \cos x_{12} + 2 \sin^{2} x_{23} - \sin x_{23} \sin{\left(x_{12} - x_{13} + x_{23} \right)} \right. \nonumber \\
 & \left. \qquad - 3 \cos x_{12} \cos x_{13} \cos x_{23} + 3 \right] \nonumber \\
 & + \frac{8 S_\mathrm{acc}}{3} \left[ \sin^{2} x_{12} - 2 \sin x_{12} \sin x_{13} - 2 \sin x_{12} \sin x_{23} - 2 \sin x_{12} \sin{\left(x_{12} - x_{13} \right)} \right. \nonumber \\
 & \left. \qquad - 2 \sin x_{12} \sin{\left(x_{12} + x_{13} \right)} \cos x_{23} + 2 \sin x_{12} \sin{\left(x_{13} + x_{23} \right)} + \sin^{2} x_{13} \right. \nonumber \\
 & \left. \qquad + 2 \sin x_{13} \sin x_{23} \cos x_{12} - 2 \sin x_{13} \sin x_{23} - 2 \sin x_{13} \sin{\left(x_{13} - x_{23} \right)} \right. \nonumber \\
 & \left. \qquad - 2 \sin x_{13} \sin{\left(x_{13} + x_{23} \right)} \cos x_{12} + \sin^{2} x_{23} + 2 \sin x_{23} \sin{\left(x_{12} - x_{23} \right)} \right. \nonumber \\
 & \left. \qquad - 2 \sin x_{23} \sin{\left(x_{12} + x_{23} \right)} \cos x_{13} - 6 \cos x_{12} \cos x_{13} \cos x_{23} + 6 \right]
\end{eqnarray} }

Unlike the Michelson-X which utilize four links from two arms, the observables from Relay configuration employ four links from three arms. When the all three arms are involved, the inequity of arms may be have trivial impact on the noise PSDs. The noise PSDs of Relay-U and Relay-T channels are expressed by distinguishing the arms, and A/E channel is specified without distinguishing the arms. Their average sensitivities are shown in right panel of Figure \ref{fig:Relay_curves}.

\subsection{Beacon and Monitor channels}

\begin{figure}[htb]
\includegraphics[width=0.46\textwidth]{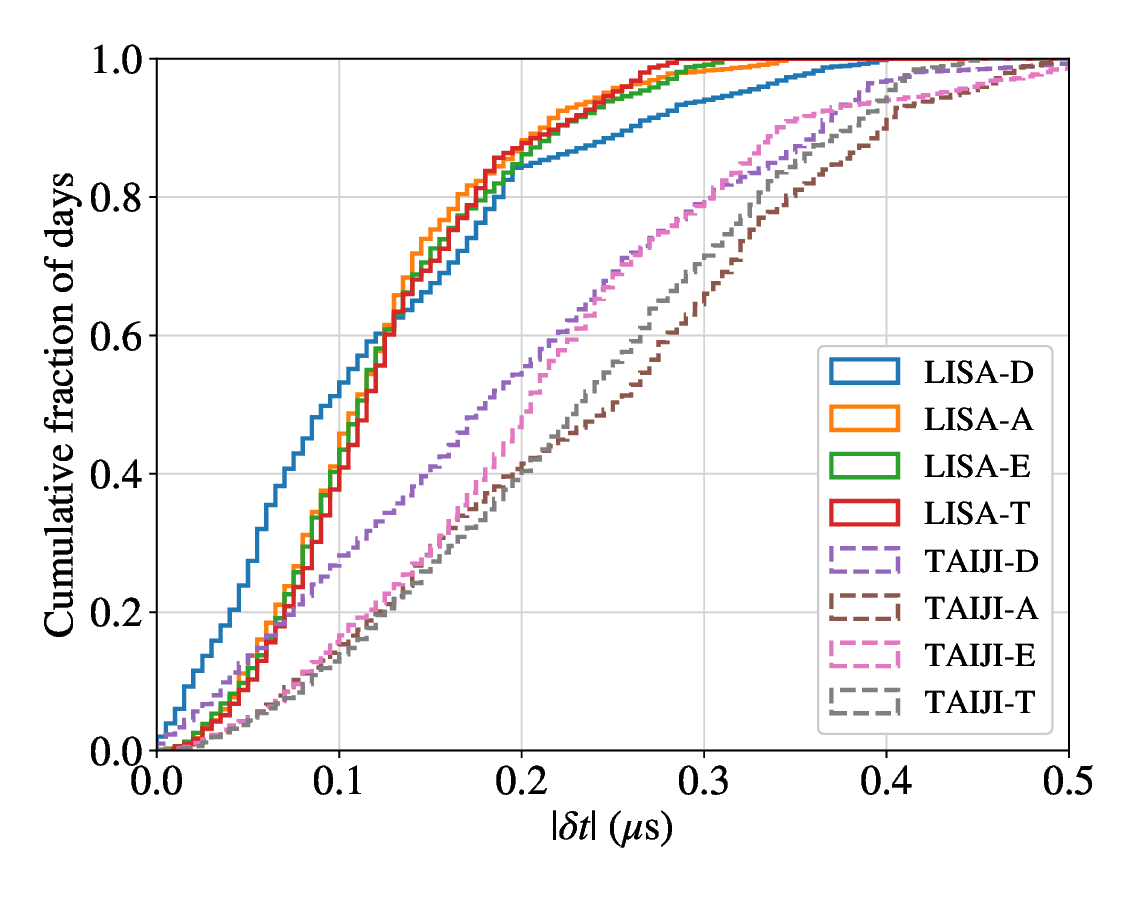}
\includegraphics[width=0.48\textwidth]{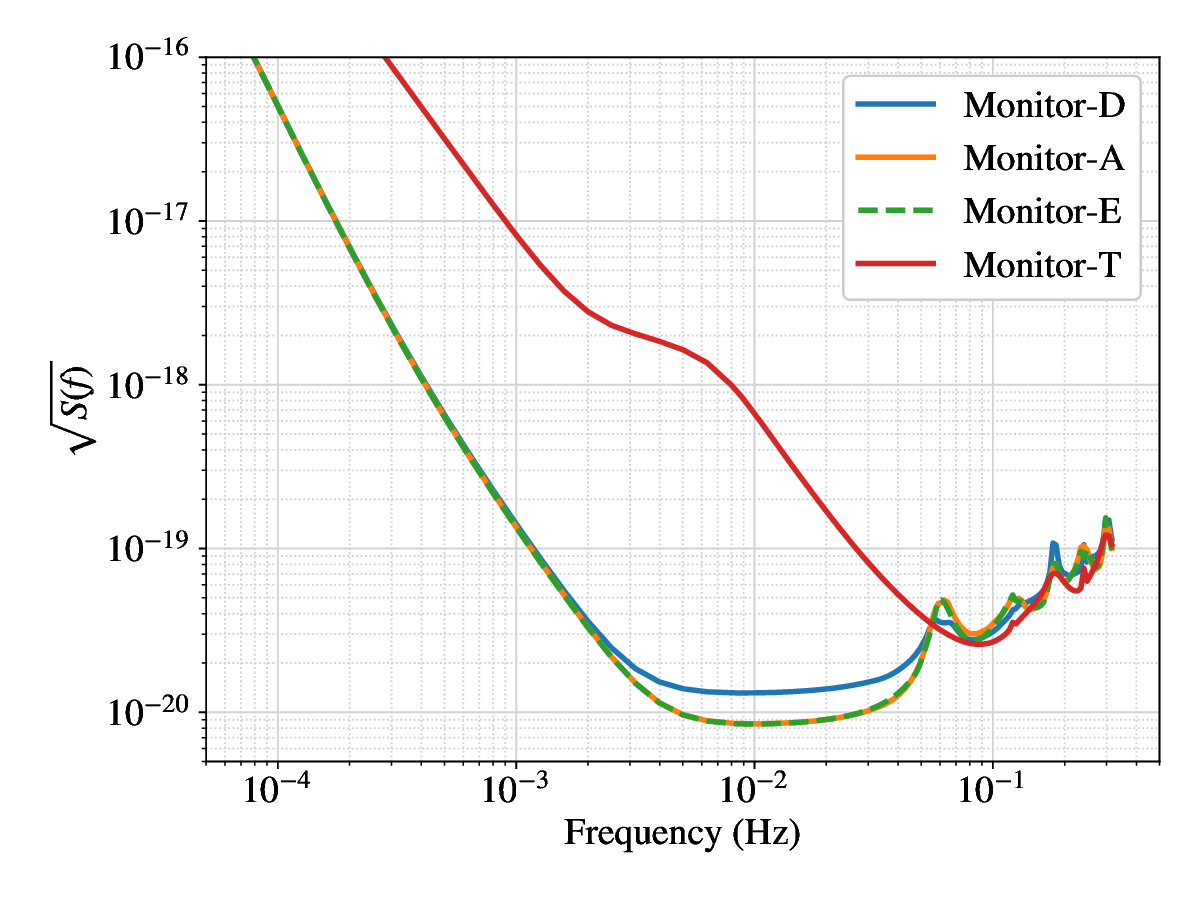}
\caption{\label{fig:Monitor_curves} The histogram of time mismatches (left panel) and average sensitivities (right panel) of Monitor-D (or Beacon-P) and the corresponding optimal channels for LISA.}
\end{figure}

The expressions of Beacon-P and Monitor-D channels are \cite{1999ApJ...527..814A,2000PhRvD..62d2002E}
\begin{eqnarray}
P =& (\mathcal{D}_{13} \eta_{32} + \mathcal{D}_{13}  \mathcal{D}_{32} \eta_{23} + \mathcal{D}_{13}  \mathcal{D}_{32}  \mathcal{D}_{23} \eta_{12} + \mathcal{D}_{12} \eta_{13} ) \nonumber \\
         & - (  \mathcal{D}_{12}  \eta_{23} + \mathcal{D}_{12}  \mathcal{D}_{23}  \eta_{32} + \mathcal{D}_{12}  \mathcal{D}_{23}  \mathcal{D}_{32} \eta_{13} + \mathcal{D}_{13} \eta_{12} ), \\
D =& (\eta_{21} + \mathcal{D}_{21}  \eta_{32} + \mathcal{D}_{21}  \mathcal{D}_{32} \eta_{23} + \mathcal{D}_{23}  \mathcal{D}_{32} \eta_{31} ) \nonumber \\
         & - (  \eta_{31} + \mathcal{D}_{31}  \eta_{23} + \mathcal{D}_{31}  \mathcal{D}_{23}  \eta_{32} + \mathcal{D}_{23}  \mathcal{D}_{32}  \eta_{21} ).
\end{eqnarray}
Their S/C layout-time delay diagrams are shown in Figure \ref{fig:X_U_P_D_diagram}. The Beacon-P could be treated as the inversed Monitor-D in the time direction. 
For the Monitor-D, the earliest time point could be at the S/C2 or S/C3, and the residual laser noise could be caused by the mismatch $\delta t_\mathrm{D}$ at $C_{23}$ or $C_{31}$. For the Beacon-P, the laser noise is related to the mismatch $\delta t_\mathrm{P}$ at laser noise $C_{12}$. From the numerical results, the mismatches from a Monitor/Beacon are equal, we deduce the reason from their diagrams. As Monitor-D plot shows, two beams (solid and dashed lines) pass by the arms $L_{23}$ and $L_{32}$ (almost) at the same time, and the effects of these links could be canceled out. Then two effective links, $L_{21}$ and $L_{31}$, are left. Similarly, the effective links for Beacon-P are $L_{12}$ and $L_{13}$. The mismatches caused by the $L_{21}$ and $L_{31}$ should be (approximately) equal to the mismatch yielded by $L_{12}$ and $L_{13}$. Compared to diagram of Michelson-X, this also make the mismatch from Monitor-D/Beacon-P is a half of the mismatch in Michelson-X, $ |\delta t_\mathrm{D} (t) | \simeq |\delta t_\mathrm{P} (t) | \simeq \frac{1}{2} | \delta t_\mathrm{X} (t) | $. 
The Monitor-D channel is selected to represent their performances.
The histogram of $\delta t$ in Monitor-D and optimal channels are plotted in the left panel of Figure \ref{fig:Monitor_curves}. 

From our previous investigation \cite{Wang:2020a}, the noise PSDs and average sensitivities from Monitor and Beacon will be also identical. 
The underlying relation between Michelson and Monitor/Beacon makes the two configurations have some common features. The inequality of arms will also affect the Monitor-T/Beacon-T channel. As the right plot of Figure \ref{fig:Monitor_curves} shows, the Monitor-T channel diverges from the equal-arm case in the lower frequency band. The PSD of secondary noise for Monitor-D and the optimal channels are described by
\begin{equation}
\eqalign{
\fl S_{\mathrm{D}} (f) =  4 S_{\mathrm{op}} \left[ 2 \sin^{2} x_{23} - \cos x_{23} \cos ( x_{12} - x_{13} ) + 1 \right] \\
+ 8 S_{\mathrm{acc}} \left[ \sin^{2} x_{23}  - 2 \cos  x_{23} \cos ( x_{12} - x_{13} ) + 2 \right], \label{eq:Sn_D}  \\
\fl S_\mathrm{A,Monitor} \simeq  S_\mathrm{E,Monitor} \simeq 4 S_\mathrm{op} ( \sin^{2}{x} - \cos{x } + 1 ) +  4 S_\mathrm{acc} ( 4 - 3 \cos{x } - \cos{3 x} ), \\
\fl S_{{\mathrm{T,Monitor}}} =  \frac{4 S_{\mathrm{op}}}{3} 
\left[ 
2 \sin^{2} x_{12} + 2 \sin^{2} x_{13} + 2 \sin^{2} x_{23} + 2 \sin x_{12} \sin x_{13} + 2 \sin x_{12} \sin x_{23}   \right. \\
 \left.  + 2 \sin x_{13} \sin x_{23} + \sin x_{12} \sin( x_{12} - x_{13} ) + \sin x_{12} \sin( x_{12} - x_{23} ) \right. \\
 \left. - \sin x_{12} \sin( x_{13} + x_{23} )  - \sin x_{13} \sin x_{23} \cos x_{12} - \sin x_{13} \sin( x_{12} - x_{13} ) \right. \\
 \left. + \sin x_{13} \sin( x_{13} - x_{23} )  - \sin x_{23} \sin( x_{12} - x_{23} ) - \sin x_{23} \sin( x_{13} - x_{23} ) \right. \\
 \left. - 3 \cos x_{12} \cos x_{13} \cos x_{23} + 3 
\right] \\
\fl \qquad \qquad +  \frac{4 S_{\mathrm{acc}}}{3} 
\left[ 
2 \sin^{2} x_{12} + 2 \sin^{2} x_{13} + 2 \sin^{2} x_{23}  + 4 \sin x_{12} \sin x_{13} + 4 \sin x_{12} \sin x_{23}  \right. \\
 \left. + 4 \sin x_{13} \sin x_{23} + 4 \sin x_{12} \sin( x_{12} - x_{13} ) + 4 \sin x_{12} \sin( x_{12} - x_{23} ) \right. \\
 \left.  - 4 \sin x_{13} \sin x_{23} \cos x_{12}   + 4 \sin x_{13} \sin( x_{13} - x_{23} ) - 4 \sin x_{12} \sin( x_{13} + x_{23} )  \right. \\
 \left.- 4 \sin x_{13} \sin( x_{12} - x_{13} ) - 4 \sin x_{23} \sin( x_{12} - x_{23} ) - 4 \sin x_{23} \sin( x_{13} - x_{23} )  \right. \\
 \left.   - 8 \sin x_{13} \sin x_{23} \cos x_{12} \cos( x_{13} - x_{23} ) - 8 \sin x_{12} \sin x_{13} \cos x_{23} \cos( x_{12} - x_{13} )  \right. \\
 \left.   - 8 \sin x_{12} \sin x_{23} \cos x_{13} \cos( x_{12} - x_{23} )  - 12 \cos x_{12} \cos x_{13} \cos x_{23} + 12
\right].
}
\end{equation}
Considering the impact of unequal-arm is trivial for Monitor-A/E, their expressions are specified without distinguishing arms.

\section{Conclusions} \label{sec:conclusions}

In this work, we examined the sensitivities of the optimal channels (A, E, and T) of the first-generation TDI configurations for LISA and TAIJI. The investigation is performed by considering the impacts of laser frequency noise, acceleration noise, and optical path noise in a time-variant unequal-arm scenario. 
We found that the T channel from Michelson and Monitor/Beacon TDI configuration is sensitive to the inequalities of the three arms both in their response functions and the noise levels in low frequencies. However, the T channel is susceptible to the laser frequency noise raised by the path mismatch of laser beams. Compared to the equal-arm case, the inequality of arm lengths could promote the detectability of the T channel. As a result, the T channel may not be an ideal null stream for noise characterization at the lower frequency band.

By using a numerical algorithm, we evaluate the path mismatches of the first-generation TDI channels. The mismatch will yield residual laser noise and affect their sensitivity especially at the characteristic frequencies. To mitigate the affects of laser noise, the stability of laser should be improved by $1-2$ orders to $0.3 - 3$ Hz/$\sqrt{\rm Hz}$ \cite{Vallisneri:2004bn}. The second-generation TDI also could be employed to overcome the laser noise \cite[and references therein]{Cornish:2003tz,Tinto:2003vj,Shaddock:2003dj,Vallisneri:2005ji,Dhurandhar:2010pd,2020arXiv200111221M}. And it has been numerically verified in our previous works \cite{Wang:2012ce,Dhurandhar:2011ik,Wang:2017aqq,Wang:2020cpq}.

For orbital optimization for a LISA-like mission, it would be harder to minimize the relative velocities between the S/C with longer arm length \cite{Wang:2017aqq}. As \Eref{eq:dL_X_approx} shows, the mismatch of the TDI beams decreases with the decrease of arm lengths and relative velocities, and the laser frequency noise will be reduced with arm length by a power index of $\gtrsim 2$. This is reflected by comparing the LISA and TAIJI result, the arm length of TAIJI is longer than LISA by 20\%, and the relative velocities between S/C of TAIJI is also larger than LISA's by 20\%. As a result, the laser frequency noise for TAIJI is higher than LISA's by $\sim$40\%. For the shorter arm LISA-like mission concept--AMIGO (nominal arm length $10^4$ km) \cite{Ni:2019amigo}, the preliminary results showed that the first-generation TDI may be sufficient to suppress the laser frequency noise under the core secondary noise and to detect the GW in the middle frequency band.

\ack
This work was supported by National Key R\&D Program of China under Grant No. 2021YFC2201903, and NSFC No. 12003059. This work made use of the High Performance Computing Resource in the Core Facility for Advanced Research Computing at Shanghai Astronomical Observatory.

\appendix

\section{The notations of the $\eta$} \label{sec:eta_notations}

\begin{eqnarray} \label{eq:eta}
  \eta_{ji} &= s_{ji} + \frac{1}{2} \left[ \tau_{ij} - \varepsilon_{ij} + \mathcal{D}_{ji} ( 2 \tau_{ji} - \varepsilon_{ji} - \tau_{jk} ) \right] \nonumber \\
  & \quad \mathrm{for} \  (2 \rightarrow 1), (3 \rightarrow 2) \ \mathrm{and} \ (1 \rightarrow 3), \\
  \eta_{ji} &= s_{ji} + \frac{1}{2} \left[ \tau_{ij} - \varepsilon_{ij}  + \mathcal{D}_{ji} ( \tau_{ji} - \varepsilon_{ji} ) + \tau_{ik} -  \tau_{ij} \right] \nonumber \\
   & \quad \mathrm{for} \  (1 \rightarrow 2), (2 \rightarrow 3)\ \mathrm{and}\ (3 \rightarrow 1).
\end{eqnarray}
The observables $s_{ji}$, $\varepsilon_{ij}$ and $\tau_{ij}$ for $j=$S/C2 $\rightarrow i=$S/C1, 3$\rightarrow$2 and 1$\rightarrow $3,
\begin{eqnarray} \label{eq:s_epsilon_tau_1}
   s_{ji} = & y^h_{ji}:h + \mathcal{D}_{ji} C_{ji}(t) - C_{ij}(t) + \mathcal{D}_{ji} N^{\rm OB}_{ji}(t) - N^{\rm OB}_{ij}(t) + n^{\rm opt}_{ij}(t), \\
   \varepsilon_{ij} = & C_{ik}(t) - C_{ij}(t) + 2 n^{\rm acc}_{ij}(t) - 2 N^{\rm OB}_{ij}(t), \\
   \tau_{ij} = & C_{ik}(t) - C_{ij}(t) ,
\end{eqnarray}
and observables $s_{ij}$, $\varepsilon_{ij}$ and $\tau_{ij}$ for $1 \rightarrow 2$, $2 \rightarrow 3$ and $3 \rightarrow 1$,
\begin{eqnarray} \label{eq:s_epsilon_tau_2}
   s_{ji} = & y^h_{ji}:h + \mathcal{D}_{ji} C_{ji}(t) - C_{ij}(t) - \mathcal{D}_{ji} N^{\rm OB}_{ji}(t) + N^{\rm OB}_{ij}(t) + n^{\rm opt}_{ij}(t), \\
   \varepsilon_{ij} = & C_{ik}(t) - C_{ij}(t) - 2 n^{\rm acc}_{ij}(t) + 2 N^{\rm OB}_{ij}(t), \\
   \tau_{ij} = & C_{ik}(t) - C_{ij}(t).
\end{eqnarray}
The symbols are specified as follows.
\begin{itemize}
\item $y^h_{ji}$ is the response function for the GW signal $h$.
\item $C_{ij}$ is the laser frequency noise of the laser on the optical bench in S/C$i$ pointing to S/C$j$.
\item $N^{\mathrm{OB}}_{ij}$ denotes the effect from displacement along $L_{ji}$ for the optical bench on S/C$i$ pointing to S/C$j$.
\item $n^{\mathrm{opt}}_{ij}$ denotes the optical path noise on the S/C$i$ pointing to the S/C$j$.
\item $n^{\mathrm{acc}}_{ij}$ is  the acceleration noise from test mass on the S/C$i$ pointing to S/C$j$.
\item $L_{ij}$ is the arm length from S/C$i$ to $j$. In this work, we assume the unequal-arm triangle configuration is static during a TDI laser beam propagation time ($L_{ij} =  L_{ji}$ and $L_{12} \neq L_{13} \neq L_{23}$) when we calculate the GW response and PSD of noises, and adopt the dynamical case when we calculate the mismatch of laser beam paths.
\end{itemize}

\bibliographystyle{iopart-num}
\bibliography{iopref}

\end{document}